\documentclass[sigconf]{acmart}
\usepackage{rotating}
\usepackage{verbatim}

\copyrightyear{2024}
\acmYear{2024}
\setcopyright{acmlicensed}\acmConference[CHI '24]{Proceedings of the CHI Conference on Human Factors in Computing Systems}{May 11--16, 2024}{Honolulu, HI, USA}
\acmBooktitle{Proceedings of the CHI Conference on Human Factors in Computing Systems (CHI '24), May 11--16, 2024, Honolulu, HI, USA}
\acmDOI{10.1145/3613904.3642764}
\acmISBN{979-8-4007-0330-0/24/05}

\begin{document}

\title[Understanding Human-AI Collaboration in Music Therapy]{Understanding Human-AI Collaboration in Music Therapy Through Co-Design with Therapists}

\author{Jingjing Sun}
\orcid{0000-0002-7005-398X}
\affiliation{%
  \institution{Institute for AI Industry Research, Tsinghua University}
  \city{Beijing}
  \country{China}}
  \email{jingjingxs08@gmail.com}

\author{Jingyi Yang}
\affiliation{%
  \institution{Institute for AI Industry Research, Tsinghua University}
  \city{Beijing}
  \country{China}}

\author{Guyue Zhou}
\affiliation{%
  \institution{Institute for AI Industry Research, Tsinghua University}
  \city{Beijing}
  \country{China}}

\author{Yucheng Jin}
\authornote{Corresponding Author}
\orcid{0000-0002-3926-7277}
\affiliation{%
 \institution{Department of Computer Science, Hong Kong Baptist University}
 \city{Hong Kong}
 \country{China}}
 \email{yuchengjin@hkbu.edu.hk}

\author{Jiangtao Gong}
\authornotemark[1]
\orcid{0000-0002-4310-1894}
\affiliation{%
  \institution{Institute for AI Industry Research, Tsinghua University}
  \city{Beijing}
  \country{China}}
\email{gongjiangtao2@gmail.com}

\renewcommand{\shortauthors}{Sun et al.}

\begin{abstract}

The rapid development of musical AI technologies has expanded the creative potential of various musical activities, ranging from music style transformation to music generation. However, little research has investigated how musical AIs can support music therapists, who urgently need new technology support. This study used a mixed method, including semi-structured interviews and a participatory design approach. By collaborating with music therapists, we explored design opportunities for musical AIs in music therapy. We presented the co-design outcomes involving the integration of musical AIs into a music therapy process, which was developed from a theoretical framework rooted in emotion-focused therapy. After that, we concluded the benefits and concerns surrounding music AIs from the perspective of music therapists. 
Based on our findings, we discussed the opportunities and design implications for applying musical AIs to music therapy. Our work offers valuable insights for developing human-AI collaborative music systems in therapy involving complex procedures and specific requirements.
\end{abstract}

%   <concept>
%       <concept_id>10003120.10003130.10011762</concept_id>
%       <concept_desc>Human-centered computing~Empirical studies in collaborative and social computing</concept_desc>
%       <concept_significance>500</concept_significance>
%       </concept>
%  </ccs2012>

% \ccsdesc[500]{Human-centered computing~Empirical studies in collaborative and social computing}

\begin{CCSXML}
<ccs2012>
   <concept>
       <concept_id>10003120.10003130.10011762</concept_id>
       <concept_desc>Human-centered computing~Empirical studies in collaborative and social computing</concept_desc>
       <concept_significance>500</concept_significance>
       </concept>
   <concept>
       <concept_id>10010405.10010444.10010446</concept_id>
       <concept_desc>Applied computing~Consumer health</concept_desc>
       <concept_significance>300</concept_significance>
       </concept>
 </ccs2012>
\end{CCSXML}

\ccsdesc[500]{Human-centered computing~Empirical studies in collaborative and social computing}
\ccsdesc[300]{Applied computing~Consumer health}

\keywords{music therapy, human-AI collaboration, co-design}


\maketitle

\section{Introduction}

Music therapy is a systematic and evidence-based clinical practice that uses a variety of musical activities~\cite{center2013music}, such as music listening, music singing, music instrument interaction, or music composition, to intervene and improve clients' mental health conditions~\cite{wheeler2015music}. In a client-centered approach, licensed music therapists lead and conduct treatment sessions primarily to assist their clients in reaching particular therapeutic objectives~\cite{bruscia1988survey}. Previous research on music therapy has found that it can alleviate people's negative emotions while treating physical diseases or be used as an effective adjuvant treatment. It has a wide range of applications, such as solving emotional problems and relieving anxiety and depression~\cite{jasemi2016effects,clark2006use,lee2013does}; or it can be used in cancer care~\cite{stanczyk2011music}, perinatal medicine~\cite{chang2008effects}, rehabilitation training~\cite{galinska2015music}, hospice care~\cite{mcconnell2017music}, and other geriatric medicine~\cite{guetin2009effect}. % Music therapy client types %

In addition to the expertise of music therapists, technology support is crucial for the complicated process that has become recognized in music therapy~\cite{baltaxe2018exploring,baglione2021understanding}. For example, an interview and survey involving 30 music therapists~\cite{baglione2021understanding} revealed their urgent technology requirements: enhanced efficiency; improved communication with clients; personalized treatment approaches; assistance in fostering clients' creative output; and the perpetual preservation of musical compositions as mementos. Thus, music technology has been adapted into music therapy practice to facilitate therapist-client communication and enhance session efficiency. 
However, those previous studies~\cite{baglione2021understanding, baltaxe2018exploring} mainly focused on how to implement digitalized music technologies, such as MIDI applications, music recording, digital music interfaces, and computer-aided music interfaces, into music therapy sessions. 

With the emergence of artificial intelligence (AI), a new area of research has arisen regarding adapting AI-related musical activities, or "musical AIs" into music therapy. Musical AIs have the potential to improve treatment efficiency and diversify treatment, aligning with both the technological practices and needs of music therapists identified in prior research~\cite{baglione2021understanding}. Current musical AIs can generate music in specific musical genres and define the instrument combinations. Deep learning models can even create melodies that manifest certain sensations~\cite{hernandez2021music}. Beyond the mere composition of melodies, other AI-based musical tasks, such as music style transfer, harmonization, tone transfer, and emotion-based music generation, are also emerging. Such technologies could have the potential to increase the effectiveness of music therapy. For example, musical AIs could help therapists better understand the relationship between music genres and human emotions and generate personalized therapeutic music~\cite{rahman2021towards, hu2020make}.

%%以下补充文献
However, despite the benefits of musical AIs, incorporating AI technologies into music therapy may require therapists to acquire new skills to use new software or electronic devices and influence their original workflow.  As well known in the Human-Computer Interaction (HCI) community, human-AI interaction and human-AI collaboration are particularly difficult to design~\cite{yang2020re}. Due to AI's output complexity and capability uncertainty, most of the early studies on musical AIs stand from the algorithm's point of view to explore the potential for combining music with technologies. Although the field of research on musical AI is still in its early stages, it is necessary to understand music therapists' opinions on current techniques and how these technologies would be integrated into or take place in music therapy. 
Specifically, we raise several research questions: \textit{What are therapists' views on the current music AI technology? How do therapists think music AI can assist in the therapy process? What concerns do therapists have about the application of music AI?} 

% Motivation and Research Gap %
To answer the above research questions, we conducted semi-structured interviews and two co-design workshops with 14 practicing music therapists specializing in treating emotional issues. During the workshops, we engaged participants who used current musical AI techniques to create treatment plans for two pre-designed cases, including the prevalence of emotional issues. After the workshops, we summarised the music therapy workflow and identified the musical AI opportunities in each stage. We offer a deep understanding of the potential benefits and concerns that musical AIs may bring to music therapy from the perspective of music therapists. Finally, we discuss our findings and conclude the design implications of future musical AI products for music therapy. 
% Thus, we contribute insights for the HCI community to develop AI products that better support music therapy and other expert AI collaborative work.

In sum, this paper makes the following contributions:
1) Empirical understanding of music therapy practices derived from semi-structured interviews and co-design workshops with 14 practicing music therapists;
2) The opportunities of leveraging Musical AIs in music therapy, as identified in co-design workshops, framed by emotion-focused therapy principles; 
3) A comprehensive analysis of musical AIs' potential benefits and concerns for music therapy, as perceived by music therapists; 
4) An in-depth analysis comparing the collaboration between expert music therapists and AI within the broader research context of human-AI collaboration; 
5) Four design implications for integrating musical AIs into music therapy.

\section{Related Work}
\subsection{HCI Research for Music Therapy}

The research on music therapy within the HCI community is diverse, with most studies following a client-centered approach to address clients' needs, such as enhancing the user experience of music therapy through user engagement~\cite{kirk2016motivating} and interactivity~\cite{lobo2019chimelight,hamidi2019sensebox,ragone2020designing,vazquez2016designing}. Researchers have proposed various technologies to enhance the interactivity of music therapy systems and engage clients more in music activities. These technologies include augmented reality~\cite{correa2009computer, lobo2019chimelight}, visualization and immersive display techniques~\cite{perez2022immersive}, and tangible user interfaces~\cite{lobo2019chimelight, hamidi2019sensebox, ragone2020designing, vazquez2016designing}.

%These studies highlight the role of HCI in facilitating client engagement and interaction within music therapy. 

Although previous research has primarily focused on client's needs, as they are evidently in need of assistance due to a lack of music knowledge, some studies have also paid attention to the needs of therapists.
To understand music therapists' needs, Baltaxe-Admony et al.~\cite{baltaxe2018exploring} explored user interface designs catering specifically to them. They found that versatility, form factors, ease of the user, standalone, and data collection are the major design factors that music therapists cared about the most when considering using interactive systems in music therapy. Moreover, Baglion et al.~\cite{baglione2021understanding} assessed the use of technology by music therapists throughout the referral, assessment, treatment, and documentation processes, employing surveys and interviews.
These studies show that therapists face many challenges when conducting therapy and require technological support. However, the current use of technology by therapists is very basic. Especially regarding the needs of therapists who want to offer personalized, efficient therapeutic services, there are currently only basic digital musical instruments, digital recording, and online streaming software available. AI technology, which has great potential to meet their needs, is still under-researched. %Specifically, what are therapists' views on the current music AI technology? How do therapists think music AI can assist in the therapy process? What concerns do therapists have about the application of music AI? 

In this study, we aim to explore the integration of musical AI into music therapy practices, seeking to enhance the HCI and AI communities' understanding of its potential benefits and applications in this field.

\subsection{Recent Advances in Musical AI Technologies}

Musical AI technologies encompass artificial intelligence systems designed for diverse music applications, including music recommendations~\cite{schedl2015music}, music generation~\cite{herremans2017functional}, genre transformation~\cite{brunner2018midi}, and tone transfer~\cite{engel2020ddsp}. 
This field has garnered significant interest within the HCI community, with numerous music-related domains poised to benefit from advancements in musical AI technology. 
Several HCI researchers have explored various AI technologies with the potential to enhance music therapy practices. This paper primarily reviews contemporary musical AI technologies that could bolster music therapy activities~\cite{hernandez2021music}. Based on their functionalities, we classify these technologies into categories such as general melody generation~\cite{Doe:2022:MuseNet,roberts2018hierarchical,hsiao2021compound}, emotion-driven music generation~\cite{hung2021emopia,grekow2021monophonic}, melody harmonization~\cite{huang2019counterpoint}, and music genre or tone transfer~\cite{engel2020ddsp,brunner2018midi}.

From the above, it is evident that musical AI technology has developed robustly; however, its application in music therapy has been minimally explored. Does the current state of musical AI offer benefits for music therapy, and what specific types of musical AI applications are required by music therapists? Addressing these questions is crucial for the advancement of technology-assisted music therapy. Consequently, this paper seeks to evaluate and implement these musical AI technologies from the music therapists' perspective, aiming to resolve these crucial questions within the therapeutic context.

\subsection{Designing Human-AI Collaboration for Therapy}

The aim of designing human-AI collaboration is to harness the unique strengths of humans and AI, ultimately improving the performance of tasks carried out by either humans or AI alone~\cite{vossing2022designing}. Several studies have shown the advantages of human-AI collaboration in art therapy and various physical health therapy practices. For instance, DeepThInk~\cite{du2024deepthink} is an AI-infused art-making system created for art therapy in collaboration with art therapists. It seeks to reduce the expertise needed for art-making and boost users' creativity. Colorbo~\cite{kim2022colorbo} is an interactive system that facilitates human-AI collaboration for therapeutic art, specifically mandala coloring, involving a complex process of coloring, color combination, and pattern analysis. Some AI-based interactive systems have been developed to aid collaborative decision-making with therapists during rehabilitation assessment~\cite{lee2020co,lee2021human}. Furthermore, given that conversations are critical in therapy, a recent study illustrated how human-AI collaboration enables more empathetic conversations in mental health support~\cite{sharma2023human}.

Current AI research in music therapy has largely focused on developing music recommendation algorithms for therapy~\cite{biswal2021can,yuan2021application} and training generative models to generate therapeutic music~\cite{hou2022ai,hu2020make,williams2020use}. However, few studies have explored human-AI collaboration in music therapy. Growing research interest aims to explore human-AI collaboration in creative music activities, including live performances~\cite{bian2023momusic, hanson2020sophiapop}, composition~\cite{louie2020novice,frid2020music}, and production~\cite{nicholls2018collaborative}. Nevertheless, most studies on user and musical AI system collaboration lack a therapeutic context.
Music therapists often have a heavy workload and are expected to be versatile~\cite{baglione2021understanding}. Music therapists need to excel not just in performance and composition but also in communication skills~\cite{pavlicevic1997music}. With the potential of hybrid intelligence to support creative music activities, the research intends to explore its role in enhancing music therapy activities like instrument playing, composing, and performing for well-being~\cite{bunt2012music}. Although human-AI collaboration in therapy shows promise, critical challenges have been identified, such as the lack of human-like empathy~\cite{carlbring2023new}, impact on the therapeutic alliance~\cite{wampold2023alliance}, and client attitudes towards AI guidance~\cite{miloff2020measuring}. 

 Most prior research in HCI related to AI systems for therapy has concentrated on aiding specific aspects or challenges within the therapy process. In contrast, our research aims to understand the collaborative dynamics between humans and AI across music therapy. Moreover, the mentioned studies ~\cite{carlbring2023new,miloff2020measuring} focus mainly on client interaction with AI, overlooking the professional therapists' collaboration with AI in therapy. Therefore, our work distinguishes itself by highlighting a unique stakeholder perspective: the collaboration between professional therapists and AI in therapy.

\section{Methodology}
To gather insights on how musical AIs would support the music therapy practice and identify the potential design opportunities and challenges, we conducted semi-structured interviews and hosted two co-design workshops with practicing music therapists. Co-design approach~\cite{spinuzzi2005methodology} is widely used in the HCI community to design and develop interactive systems involving special groups of people, such as domain experts. 
The collaboration aims to facilitate and ensure that the final design aligns with the users' needs, preferences, and expectations. 
In our study, adopting the participatory design method allows us to deepen our understanding of how music therapy can incorporate musical AIs and facilitate feedback collection during each implementation process. Another reason to host a co-design workshop is that music therapists come from diverse backgrounds and follow various therapy methods. From our interview session, we cannot agree on using musical AI to implement it in a wider music therapy context. 
Thus, after the interview session, we conducted a further co-design session.

\subsection{Overall Research Protocol}
We conducted semi-structured interviews and developed two virtual design workshops to explore and understand how musical AIs would assist music therapists during the treatment process. 
Semi-structured interviews were conducted with six music therapists (see Table~\ref{MT_info}, I1-I6) to understand their basic workflow, typical therapy cases, and current pain points. The detailed interview guidelines are in Appendix C.
We first conducted a group interview with three participants who graduated from the same institute, then interviewed the rest separately. 
One of the one-to-one interviews was conducted in person, while others were conducted online via the video conferencing tool Tencent Meeting~\cite{Doe:2022:Tencentmeeting}.

Two design workshops were also conducted remotely following the same procedure with eight music therapists (see Table~\ref{MT_info}, P1-P8, four in each workshop). 
We used an online collaborative tool, Teamind~\cite{Doe:2022:Teamind} (see Appendix E, Fig.~6 and Fig.~7), to create a shared control design space to facilitate the participants to write or draw their ideas.
To avoid unnecessary difficulties encountered in using the online design tool, four researchers (the stance of the research team is reported in Section~\ref{sec:Research_Team}) took on the role of facilitators in each workshop to support participants in managing the board and putting ideas on the shared canvas. We also inform participants to prepare paper and pens in advance.

 \subsection{Procedures}

\subsubsection{Interview Procedure}
Semi-structured interviews lasted from one and a half to two hours. Before the interviews, we developed a semi-structured interview guide through an iterative process involving all research team members. The questions started by asking participants to share their current practices and work progress and then asked about their current issues with their current patient types.

 \subsubsection{Co-design Workshop Procedure}       

The whole workshop is designed to last four hours. Three days before the workshop, we sent prepared AI-generated music material (see Sections~\ref{sec:AIlearning}) to each participant and suggested that the learning process would take one or two hours. We encouraged participants to interact with the models and listen to the generated music before participating in the workshop session. However, our main objective was for participants to understand the functionalities and understand what musical AIs can do. Therefore, we did not require participants to test all the systems. However, we aim to understand the participants' natural reactions and preferences towards musical AIs after they have acquired knowledge from the learning material. Additionally, we are interested in understanding the participants' learning experience. Therefore, we asked them to fill out an interview survey to gather their subjective feelings about the learning process and receive comments on musical AIs. All participants expressed positive sentiments towards the learning process and did not report experiencing cognitive overload. 

During the workshop, we first provided a 20-minute introduction to introduce the study background, recall the AI-generated music materials, and provide the pre-designed cases (see Section~\ref{sec:case}). Next, we asked participants to use 15 minutes to work independently on an online collaborative whiteboard to brainstorm possible treatment solutions for the two cases using musical AI techniques. To facilitate the following discussion, we set up a 15-minute ice-breaking game that allows participants to become familiar with each other. Each participant wrote down their hometown, favorite song, and most memorable experiences in the activity. Then, we went to a 60-minute personal design and 90-minute co-design sessions. In the first workshop, we divided the four participants into two random groups to discuss their designs and prepare for the group presentation. However, based on the feedback from the first workshop that participants were more willing to discuss in the group, we directly moved to the discussion session in the second workshop. In the last 20 minutes, we asked participants to report their design outcomes to all the participants and make comments based on them. After the design workshop, we invited participants to participate in individual 20-minute semi-structured interviews where we asked about their experience during the workshops and subjective feelings about musical AIs. 

\subsubsection{Ethical Considerations and Others}
Before initiating the study, participants were encouraged to express their willingness to participate in all prescribed activities, understanding that they could withdraw from any task at any time if they encountered any discomfort. In our study, all participants reported a positive willingness to participate in all activities. All of the studies were conducted within a Chinese language context, and the quotes were translated into English afterward.

\subsection{Materials}
 
\subsubsection{AI-music learning materials.}\label{sec:AIlearning}  We examined current state-of-the-art musical AI technologies and selected several of the most representative works as the AI materials for our study~\cite{hernandez2021music}. 

We categorized the selected models based on their functionalities and divided them into five categories: (1), Technique A: General Melody Generation (\textbf{Ge-Gen}); (2), Technique B: Emotion-based Music Generation (\textbf{Emo-Gen}); (3), Technique C: Melody Harmonization (\textbf{Melo-Har}); (4), Technique D: Music Genre Transfer (\textbf{Genre-Tran}); and (5), Technique E: Tone Transfer (\textbf{Tone-Tran}). The description of each category and the corresponding sample models can be found in Table~\ref{AI materials}.

To adequately prepare workshop participants to utilize musical AI technologies, we created educational materials encompassing a comprehensive overview of musical AI technologies. Each algorithm's capabilities are detailed within specific categories, accompanied by corresponding deep-learning models. Furthermore, we supplied textual introductions, audio samples, online video tutorials, and live demonstrations for each model. Additionally, a condensed version of the instruction was developed to facilitate quick recollection of terms and enable easy use during the workshops, featuring concise introductions of each music algorithm function and a limited selection of audio samples.  
In the detailed version of the learning materials, we included four open-source musical AI systems: Melody Mixer~\footnote{\url{https://experiments.withgoogle.com/ai/melody-mixer/view/}}, Beat Blender~\footnote{\url{https://experiments.withgoogle.com/ai/beat-blender/view/}},~DeepBach~\footnote{\url{https://www.google.com/doodles/celebrating-johann-sebastian-bach}}, and COCOCO~\footnote{\url{https://pair-code.github.io/cococo/}}. Due to not all current musical AI research, including interface design, other music materials were directly pre-generated by our researchers. We collected the participants' interest in five kinds of musical AIs through a pre-workshop survey (N = 6; two participants didn't finish our questionnaire; the survey details are in Appendix D). The results show that \textbf{Ge-Gen} received a 16.67\% score while the other four techniques all received a 66.67\% score. We did not specifically gather information on why \textbf{Ge-Gen} received a lower score than other categories. However, based on our participants' responses to the survey, P4 mentioned that if \textbf{Ge-Gen} is only focused on generating various types and styles of music without a clear link to the purpose of therapy, it may not be effective. Therapy is a nuanced and intricate process that demands substantial learning. We can speculate that participants questioned whether the generated music would effectively balance personalized interest and therapeutic function. This might be caused by the low satisfaction level of the \textbf{Ge-Gen} technique.% We encouraged participants to interact with the models and listen to the generated music before participating in the workshop session. However, our main objective was for participants to understand the functionalities and gain a basic understanding of what musical AIs can do. Therefore, we did not require participants to test all the systems. Instead, we asked them to fill out an interview survey to gather their subjective feelings about the learning process and receive comments on musical AIs. All participants expressed positive sentiments towards the learning process and did not report experiencing cognitive overload.
 
\begin{table*}
\centering
\caption{Five categories of AI materials}
\label{AI materials}
\begin{tabular}{lp{2cm}p{6cm}p{3cm}} 
\hline
        \textbf{Techniques ID}  & \textbf{AI Techniques}         & \textbf{Main Functions}
            & \textbf{Sample Models}                                                                    \\ 
\hline
\textbf{Ge-Gen} & General Melody Generation      & \begin{tabular}[c]{@{}p{6cm}@{}}Generate music from scratch: randomly generate melodies from scratch\\ Generate by extending intro music: randomly generate melodies that extend given intro music\\ Create transition: Create a harmonic transition between two separate melodies by performing music interpolation\end{tabular} & \begin{tabular}[c]{@{}l@{}}MuseNet~\cite{Doe:2022:MuseNet} \\ MusicVAE~\cite{roberts2018hierarchical}\\ Compound Word\\ Transformer~\cite{hsiao2021compound}\end{tabular}  \\ 
\hline
\textbf{Emo-Gen} & Emotion-based Music Generation & Emotion-based generation: produce music pieces with disparate emotions that may be characterized using Russell's circumplex model.                                                                                                                                                                              & \begin{tabular}[c]{@{}p{6cm}@{}}EMOPIA~\cite{hung2021emopia} \\ Conditional VAE~\cite{grekow2021monophonic}\end{tabular}                         \\ 
\hline
\textbf{Melo-Har} & Melody Harmonization           & Harmonize prime melody: produce a harmonized music piece when given a prime melody                                                                                 & Coconet~\cite{huang2019counterpoint}                                                                                   \\ 
\hline
\textbf{Genre-Tran} & Music Genre Transfer           & \begin{tabular}[c]{@{}p{6cm}@{}}
Transfer genres: adapt an existing music piece to a different music genre (e.g., a jazz style to a classical style)
\\Create song mixtures: create mixtures of multiple distinctive songs\end{tabular}                                                                                & MIDI-VAE~\cite{brunner2018midi}                                                                                  \\ 
\hline
\textbf{Tone-Tran} & Tone Transfer                  & Transfer tone: transform sounds into musical instruments (e.g., transforming the sound of birds chirping into violin performances)                                                                                                                                                                               & DDSP~\cite{engel2020ddsp}                                                                                      \\
\hline
\end{tabular}
\end{table*}

\subsubsection{Pre-designed cases.}~\label{sec:case} We noticed that music therapists have varying expertise and treatment methods. Meanwhile, as their treatment is client-centered, clients' characteristics, medical problems, and responses can greatly affect what kind of treatment and techniques music therapists use. As such, we used a persona-based method to facilitate the brainstorming amongst therapists and allow them to focus more on using AI-generated music.  We designed two cases related to emotional issues related to depression, anxiety, and stress, which are the top 25\% percent of increased mental health problems triggered by the COVID-19 pandemic~\cite{Doe:2022:Website}. We diversified the two cases based on gender, age, personality, music background, and emotional problems to reflect the different client groups. This approach would allow participants to have more design possibilities (see detailed case descriptions in Appendix A.1). 
\begin{enumerate}
\item Case 1 describes a female college student experiencing depressed emotions due to academic stress. She exhibits symptoms of depression, including a persistently sad mood, diminished interest in daily activities, inappropriate feelings of guilt, and sleeping issues. She is an introverted individual who finds it difficult to express herself. In her childhood, she used to play the piano.
\item Case 2 is a middle-aged male recently laid off. He shows typical symptoms of restlessness, irritability, and anxiety. The increased pressure led him to start drinking, which further caused emotional conflicts with his family members. He used to be outgoing and enjoy going to live concerts with friends. He has no previous musical instrument-playing experience. 
\end{enumerate}

%  \end{itemize}
%  \begin{figure}[h]
% \centering
% \includegraphics[width=1\columnwidth]{img/cases.png}
%   \caption{\textbf{Pre-designed Cases} }
%     \label{fig:user_cases}
% \end{figure}

 \subsection{Participants}
 %%Add recruitment process and split with interview people?
Due to the various techniques that music therapy includes, we require participants to have experience in treating clients with anxiety, stress, and depression. We recruited participants via snow sampling and online advertising on WeChat. Overall, 14 participants (1 male,13 female, aged 25 to 40) were hired. All of the participants are Chinese. However, some of them have working experience in different countries (see the following details in Table ~\ref{MT_info}). We have conducted inquiries regarding music therapist certification in different countries. In the United States, individuals must obtain Music Therapy Board Certification (MTC-C). In Canada, the certification of Music Therapist (MTA) can be obtained through the Canadian Association of Music Therapists. Each music therapist who participated in interviews and the co-design workshop was compensated based on their professional hourly rate, which amounted to approximately \$200 per hour. 

\subsection{Research Team and Stance}\label{sec:Research_Team}
Our research team included people with backgrounds in human-computer interaction, music therapy research, design, and AI algorithm research and development. One researcher is a senior Ph.D. candidate with experience conducting actual small-scale music therapy, which helped us develop workshop pre-design cases. Each group in our co-design sessions consisted of at least one designer and one AI algorithm researcher.

\subsection{Data Analysis}

\begin{table*}
\footnotesize
\caption{Demographics of Music Therapist in This Study}
\label{MT_info}
\begin{tabular}{lllllp{4cm}l}
\hline
\textbf{ID} & \textbf{\begin{tabular}[c]{@{}l@{}}Working\\ Years\end{tabular}} & \textbf{\begin{tabular}[c]{@{}l@{}}Country\end{tabular}} & \textbf{\begin{tabular}[c]{@{}l@{}}Certification\end{tabular}} & \textbf{\begin{tabular}[c]{@{}l@{}}Applied \\ Theories\end{tabular}}                        
& \textbf{Clients}                            & \textbf{\begin{tabular}[c]{@{}l@{}}Knowledge\\of AI\end{tabular}} \\ \hline
\\
I1         & 5$\sim$10                                                        & \begin{tabular}[c]{@{}l@{}}China\end{tabular}              & \begin{tabular}[c]{@{}l@{}}MTA; \\ Certified \\ in China\end{tabular}              & \begin{tabular}[c]{@{}l@{}}Humanistic \\ psychology\end{tabular}
& \begin{tabular}[c]{@{}p{4cm}@{}}Humanistic psychology
Currently teaching in college
\end{tabular}                          & \begin{tabular}[c]{@{}l@{}}Not know \\ at all\end{tabular}    
\vspace{0.1cm}
\\
I2         & 3$\sim$5                                                        & \begin{tabular}[c]{@{}l@{}}Malaysia\end{tabular}              & \begin{tabular}[c]{@{}l@{}} MTA\end{tabular}              & \begin{tabular}[c]{@{}l@{}}Humanistic \\ psychology; \\ Psychodynamic \\ psychology\end{tabular}
& \begin{tabular}[c]{@{}p{4cm}@{}}Children with special needs; Rehabilitated patients; Elderly with emotion issues\end{tabular}                          & \begin{tabular}[c]{@{}l@{}}Not know \\ at all\end{tabular}    
\vspace{0.1cm}
\\
I3        & 3$\sim$5                                                        & \begin{tabular}[c]{@{}l@{}}China\end{tabular}              & \begin{tabular}[c]{@{}l@{}}Certified \\ in China\end{tabular}              & \begin{tabular}[c]{@{}l@{}}Humanistic \\ psychology\end{tabular}
& \begin{tabular}[c]{@{}p{4cm}@{}}Elderly with emotion issues\end{tabular}                          & \begin{tabular}[c]{@{}l@{}}Have heard \\ it but never\\ used it\end{tabular}    
\vspace{0.1cm}
\\
I4         & 5$\sim$10                                                        & \begin{tabular}[c]{@{}l@{}}China\end{tabular}              & \begin{tabular}[c]{@{}l@{}}Certified \\ in China\end{tabular}              & \begin{tabular}[c]{@{}l@{}}CBT; \\ DBT\end{tabular}
& \begin{tabular}[c]{@{}p{4cm}@{}}Adults with obsessive-compulsive disorder; Depression; Persecutory delusion; sleeping disorder; Eating disorder\end{tabular}                          & \begin{tabular}[c]{@{}l@{}}Not know \\ at all\end{tabular}    
\vspace{0.1cm}
\\
I5         & 5$\sim$10                                                        & \begin{tabular}[c]{@{}l@{}}China/\\ Canada\end{tabular}              & \begin{tabular}[c]{@{}l@{}}MTA; \\ MT-BC\end{tabular}              & \begin{tabular}[c]{@{}l@{}}Humanistic \\ psychology; \\ Psychodynamic \\ psychology\end{tabular}
& \begin{tabular}[c]{@{}p{4cm}@{}}People with depression and anxiety; Children with special needs; elderlies; People with disabilities\end{tabular}                          & \begin{tabular}[c]{@{}l@{}}Not know \\ at all
\end{tabular}    
\vspace{0.1cm}
\\

I6         & 5$\sim$10                                                        & \begin{tabular}[c]{@{}l@{}}China /\\ United State\end{tabular}              & \begin{tabular}[c]{@{}l@{}}MTA; \\ MT-BC\end{tabular}              & \begin{tabular}[c]{@{}l@{}}Humanistic \\ psychology; \\ Psychodynamic \\ psychology\end{tabular}
& \begin{tabular}[c]{@{}p{4cm}@{}}Children with special needs; Alzheimer's; Adults with depression; Schizophrenia\end{tabular}                          & \begin{tabular}[c]{@{}l@{}}Have heard \\ it but never\\ used it\end{tabular}    
\vspace{0.1cm}
\\
 
\midrule

P1         & 5$\sim$10                                                        & China                                                                & MT-BC                                                              & \begin{tabular}[c]{@{}l@{}}Humanistic \\ psychology; \\ CBT\end{tabular}
& \begin{tabular}[l]{@{}p{4cm}@{}}Adults with anxiety and depression symptoms; Teenagers and adults with diagnosed depression and anxiety\end{tabular}                    & \begin{tabular}[c]{@{}l@{}}Not know \\ at all\end{tabular}                   
\vspace{0.1cm}
\\
P2         & 5$\sim$10                                                        & China                                                                & \begin{tabular}[c]{@{}l@{}}Certified \\ in China\end{tabular}      & \begin{tabular}[c]{@{}l@{}}Humanistic \\ psychology\end{tabular}
& \begin{tabular}[c]{@{}p{4cm}@{}}General psychological issues, including depression, and anxiety\end{tabular}                                                                   & \begin{tabular}[c]{@{}l@{}}Not know \\ at all\end{tabular}

\vspace{0.1cm}
\\
P3         & 5$\sim$10                                                        & \begin{tabular}[c]{@{}l@{}}China /\\ Canada\end{tabular}              & \begin{tabular}[c]{@{}l@{}}MTA; \\ MT-BC\end{tabular}              & \begin{tabular}[c]{@{}l@{}}Humanistic \\ psychology; \\ Psychodynamic \\ psychology\end{tabular}
& \begin{tabular}[c]{@{}p{4cm}@{}}People with depression and anxiety; Children with special needs; elderlies; People with disabilities\end{tabular}                          & \begin{tabular}[c]{@{}l@{}}Have heard \\ it but never\\ used it\end{tabular}    
\vspace{0.1cm}
\\
P4         & 3$\sim$5                                                         & China                                                                & \begin{tabular}[c]{@{}l@{}}Certified \\ in China\end{tabular}      & \begin{tabular}[c]{@{}l@{}}Psychodynamic \\ psychology\end{tabular}
& \begin{tabular}[c]{@{}p{4cm}@{}}Maternal depression; Teenagers with depression, schizophrenia, personality disorder\end{tabular}                                           & \begin{tabular}[c]{@{}l@{}}Have heard \\ it but never\\ used it\end{tabular}    
\vspace{0.1cm}
\\
P5         & 5$\sim$10                                                        & Singapore                                                            & \begin{tabular}[c]{@{}l@{}}Certified \\ in China\end{tabular}       & \begin{tabular}[c]{@{}l@{}}Humanistic \\ psychology\end{tabular}                                        

& \begin{tabular}[c]{@{}p{4cm}@{}}Adults with depression, anxiety, bipolar disorder, schizophrenia\end{tabular}                                                                 & \begin{tabular}[c]{@{}l@{}}Have heard \\ it but never\\ used it\end{tabular}    
\vspace{0.1cm}

\\
P6         & 5$\sim$10                                                        & \begin{tabular}[c]{@{}l@{}}United \\ States\\ / China\end{tabular}   & \begin{tabular}[c]{@{}l@{}}MTA; \\ MT-BC\end{tabular}              & \begin{tabular}[c]{@{}l@{}}Humanistic \\ psychology; \\ Psychodynamic\\ psychology\end{tabular} 
& \begin{tabular}[c]{@{}p{4cm}@{}}Adults with cancer, schizophrenia, emotional disorder, depression, anxiety, and stress; special care baby unit\end{tabular}             & \begin{tabular}[c]{@{}l@{}}Have heard \\ it but never\\ used it\end{tabular}    
\vspace{0.1cm}
\\
P7         & 10+                                                              & China                                                                & \begin{tabular}[c]{@{}l@{}}Certified \\ in China\end{tabular}       & \begin{tabular}[c]{@{}l@{}}Humanistic \\ psychology; \\ CBT\end{tabular}
& \begin{tabular}[c]{@{}p{4cm}@{}}Adult with emotional issues; oncology clients; autism; Alzheimer's; prison inmates; newborns\end{tabular}                                  & \begin{tabular}[c]{@{}l@{}}Have heard \\ it but never\\ used it\end{tabular}  \vspace{0.1cm}  
\\
P8         & 3$\sim$5                                                         & \begin{tabular}[c]{@{}l@{}}United \\ States\end{tabular}             & MT-BC                                                              & \begin{tabular}[c]{@{}l@{}}Humanistic \\ psychology; \\ Psychodynamic \\ psychology\end{tabular}
& \begin{tabular}[c]{@{}p{4cm}@{}}Adults with depression, anxiety, phobia, psychological disorders, schizophrenia, bipolar disorder, and split personality\end{tabular} & \begin{tabular}[c]{@{}l@{}}Have heard\\ it but never\\ used it\end{tabular}     \\
\midrule
\end{tabular}
\end{table*}

We collected data from all interviews and co-design workshops, including video and audio recordings. The audio recording materials were transcribed verbatim by two researchers. Thematic analysis, which focused on the semantic interpretation of the data, was used to conceptualize themes from the co-design workshops and the interviews and further followed with an inductive open coding approach. The two researchers carried out initial coding, and as the analysis progressed, the conceptualization themes were discussed among the research group members.

\section{Findings}

\subsection{The Co-Design Music Therapy Treatment Plan}

During co-design workshops, all of our participant music therapists designed and shared their therapy ideas in Case 1 and Case 2. 
Following multiple rounds of discussion, they agreed on the music therapy plans. We concluded the typical music therapy workflow they utilized as a working procedure and outlined their therapy plan accordingly.

%The co-design workshop not only provided a working framework for musical AI Therapy but also further improved the design ideas of it as well.
% Figure~\ref{fig:stages_therapy}) summarizes the activities of two co-design workshops, where the therapists planned the work of the two virtual clients within the music therapy framework they have shown. Including the music therapy methods and AI algorithm used in each intervention stage:
\subsubsection{Typical Workflow of Music Therapy}
\begin{figure*}[h]
\centering
  \includegraphics[width=\textwidth]{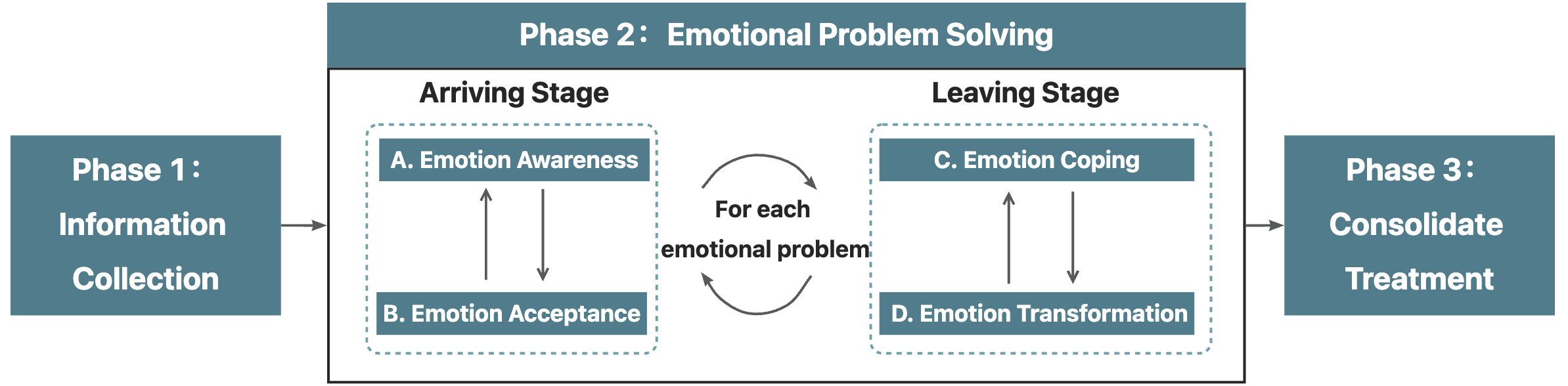}
  \caption{\textbf{Typical Workflow of Music Therapy} }
    \label{fig:therapy_process}
    \Description{Figure 1 illustrates the workflow of music therapy for emotional problems. This workflow comprises three phases: the initial phase is information collection, and the concluding phase is treatment consolidation. The pivotal treatment phase falls between the first and third phases, encompassing several elements, including emotional awareness, acceptance, coping, and transformation. The first two stages serve as the identification stage (arriving) for emotional problems, while the latter two constitute the resolution stage (leaving) for emotional problems.}
\end{figure*}

All participants (14/14) agreed that, up to now, music therapy has developed into different therapeutic genres and therapeutic techniques. However, there is no single unified working workflow for music therapy. To investigate potential musical AI design opportunities in music therapy, we adhere to a standard music therapy workflow derived from discussions with the participating therapists. This workflow also aligns with general psychotherapy practices. %From the information provided by our participant music therapists, music therapists see therapy as "identifying and solving problems."---which means that, in the process of treatment, the therapist will first help the client to recognize and understand their problems or needs and try to solve these problems in the subsequent process. 
% Thus, the process is divided into three phases which include identifying the problem, solving the problem, and consolidating the effect (similar with~\cite{mearns2013person}). 
Throughout our design workshop, the workflow of music therapy around emotional issues can be divided into three phases: 1) Information Collection, 2) Emotional Problem-Solving, and 3) Treatment Consolidation (see Figure~\ref{fig:therapy_process}). In the phase of Information Collection, the therapist gathers information about the client's musical preferences, history, and emotional concerns. During the phase of Emotional Problem-Solving, the therapist uses music to address and work through the client's emotional challenges and issues. Treatment Consolidation refers to integrating the progress made in therapy and utilizing music to maintain emotional well-being and growth.
Our study mainly focused on using musical AIs in the Emotional Problem-Solving phase. Emotional Problem-solving is similar to emotion-focused therapy, containing two core stages, emotion arriving and emotion leaving~\cite{greenberg2004emotion}. At the \textbf{arriving stage}, \underline{emotion awareness} is primarily designed to help clients \emph{experience} and \emph{express} their emotions, associated with \underline{emotion acceptance} that helps clients \emph{understand} and \emph{accept} their current emotions. Once the client is fully aware of and accepts their emotions, the focus shifts to the \textbf{leaving stage}. At this stage, the therapy session centers on coping with the client's emotional issues (\underline{emotion coping}) and transforming them into the desired emotion (\underline{emotion transformation})~\cite{greenberg2004emotion}. During emotion coping, therapists will assist clients in \emph{coping and addressing current emotional issues} first. Once relief is achieved for the current emotional issues, therapists will help clients further \emph{transform their emotions}, facilitating the emergence of new emotional states. Since emotions are dynamic, the Emotional Problem-Solving phase involves an iterative process, potentially revisiting stages until the client successfully achieves the target emotion. Subsequently, the therapy transitions to the consolidation phase.

%With musical AI technologies and two pre-designed cases as stimuli, the participant music therapists brainstormed with our researchers, and the therapists developed treatment plans by integrating musical AI techniques. In the following sections, we provide a detailed explanation of the music therapy framework and summarize therapy plans for each case.
%%Still need to shorten this quot%%
%Greenberg~\cite{greenberg2004emotion} explains: ``A major premise is that one cannot leave A place until one has arrived at it.'' It means that to solve an emotional problem, we need to present the problem first, and then think about solutions. As such,  "arriving" is considered the first phase, which is focusing on the "awareness and acceptance of an emotion". The second phase is called "Leaving", which focuses on "emotion utilization" or "transformation." This stage involves moving on or transforming core feelings~\cite{greenberg2004emotion}. These two phases correspond to problem identification and problem resolution. We also found a similar workflow for music therapy in the present study.

\begin{figure*}[h]
\centering
\includegraphics[width=\textwidth]{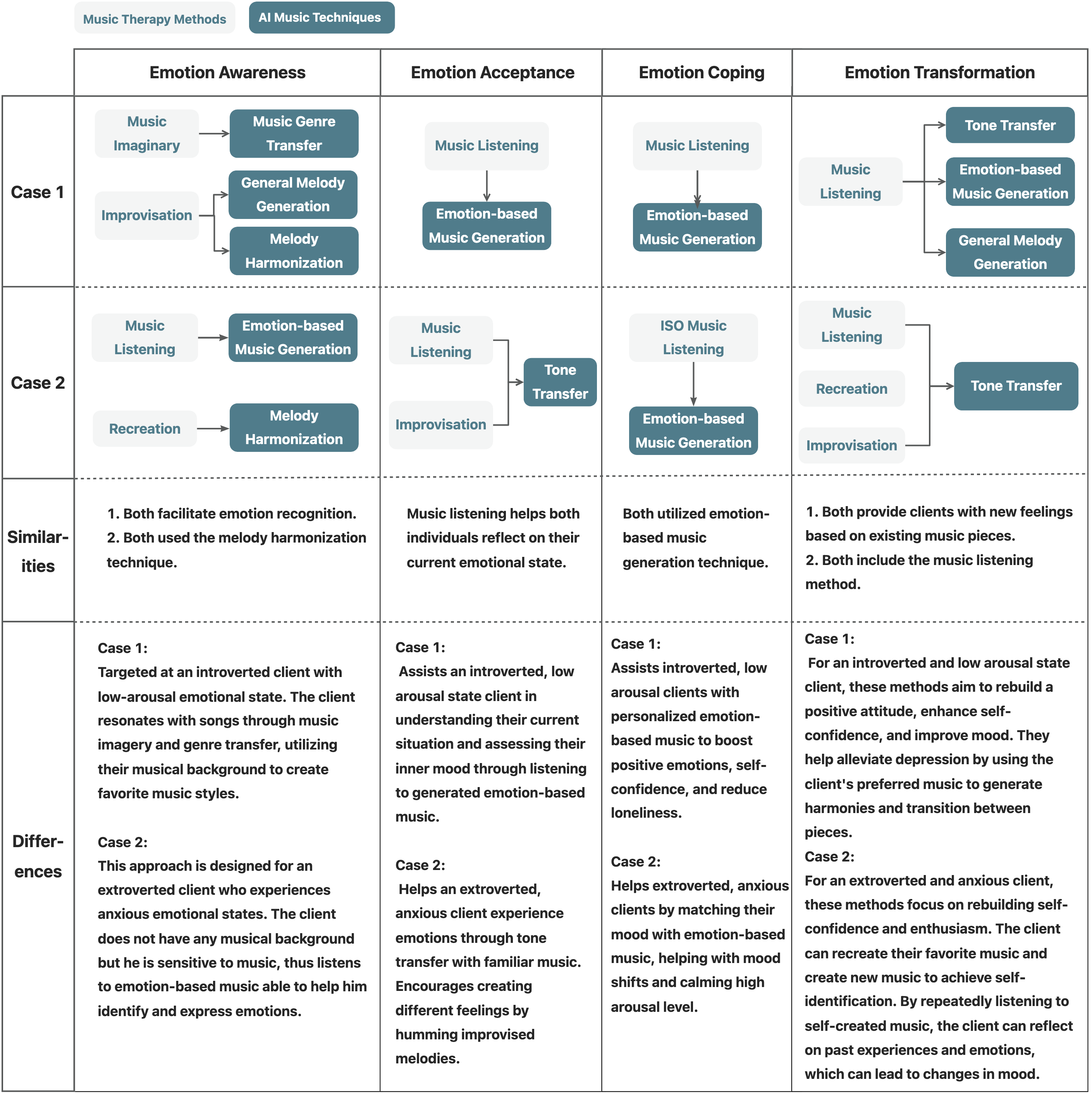}
  \caption{\textbf{Different stages of emotional problem-solving in music therapy in two cases: matching of music therapy methods and AI algorithms} }
    \label{fig:stages_therapy}
    \Description{Figure 2 illustrates the alignment between Music Therapy techniques and AI algorithms. Following the four stages of emotional problem-solving, we devised intervention strategies for Case 1 and Case 2. These interventions were paired with the appropriate AI algorithms to facilitate therapy.
}    
\end{figure*}

%At the beginning of the study, we tried to summarize a model for the application of music therapy by interviewing different therapists. 
%Based on the workflow of Emotion-focused therapy and the summary of our later data, we summarize a working workflow for music therapy that will help us to clarify the use of AI techniques in the process of music therapy.

%Similar to Emotion-focused therapy, we found that the workflow (see Figure~\ref{fig:therapy_process}) of music therapy can also be divided into two major phases: problem presentation and problem resolution. And the two phases consist of four stages: Emotion Awareness; Emotion Acceptance; Emotion Coping, and Emotion Transformation. This is the main part of the treatment. And at both ends, there is a preparation stage before treatment for collecting information and an end-stage for consolidating the effect of treatment. Here we will only try to focus on the main part of the treatment. We will describe the four stages of treatment as follows.

\subsubsection{Music Therapy Treatment Plan with Musical AIs}
Using musical AI technologies and two pre-designed cases as prompts, the therapists collaborating with our researchers engaged in brainstorming sessions and formulated treatment plans incorporating musical AI techniques. Following our derived workflow of music therapy (See Figure.~\ref{fig:therapy_process}), we summarized their therapy plans. We described how therapists treated clients with two different treatment cases, including the music therapy methods and musical AI techniques used at each treatment stage as follows (See Figure.~\ref{fig:stages_therapy} ).

\textit{(1) Emotional Awareness.} Music therapists often first help clients understand their own emotions. The first stage is dedicated to emotional awareness, in which the therapist will guide the client to evoke, experience, and express their negative thoughts and feelings.
At this stage, both Case 1 and Case 2 focused on facilitating emotion recognition, which helps clients become aware of their current emotions. However, participants mentioned that different client personalities might require slightly different treatment strategies and different techniques can be employed in diverse ways. For instance, in Case 1, as an introverted client and in a low-arousal emotional state, participants suggested using \textbf{Genre-Tran} in music imagery to assist in resonating emotions by transforming to different music styles. This helps stimulate the client's imagination and enhance emotional awareness. %Introverts and clients who tend to have depression signs (Case 1) will mostly focus on improvisation to support emotional expression. On the other hand, extroverted clients (Case 2) would be more focused on encouraging participation in music activities. 

\begin{quote}
\textit{``Let the client choose a favorite song and generate emotional resonance. Then, through the changing function of the music style, the client's imagination is promoted, and the emotions expressed by the client may be different from those expected by the therapist.''} (P2)
\end{quote}

What's more, for clients similar to Case 1, who may struggle to express their emotions and remain unaware of their own feelings, participants suggested using the improvisation method to facilitate the expression process~\cite{macdonald2014musical}. Since the client has a previous musical background, \textbf{Ge-Gen} and \textbf{Melo-Har} have been proposed to streamline the creative process and assist clients in automatically generating accompanying music.

% In the use of music imaginary, \textbf{music genre transfer} being used to transfer between music genres would assist in changing to different music styles to stimulate the client's imagination and complete the emotional awareness. 

Additionally, music listening is employed to aid clients in developing awareness of their emotions. Therapists thought clients could gradually recognize their unresolved emotions through music listening. The key is to use emotional music that aligns with the client's emotions, facilitating their immersion in and awareness of their emotional state. However, novice therapists may struggle to create music that matches their clients' current emotions. Therefore, employing \textbf{Emo-Gen} would be better to emphasize with clients.

\begin{quote}
\textit{``The algorithm for generating emotional melodies can help the client understand and define his current emotions or those that have occurred more frequently recently. This can be a tool to help him understand his emotions.''} (P4)
\end{quote}

Moreover, the recreation method has been mentioned to help clients become aware of their emotions and carry out subsequent interventions more effectively in Case 2. It is especially crucial for clients with anxiety to be cognizant of their emotions~\cite{watson2017emotion}. %However, clients with extroverted personalities and temper issues tend to find it easier to make emotional expressions and participate in activities. 
The recreation method would extend their level of participation and help them recognize their emotions. The \textbf{Melo-Har} technique is mentioned to support the recreation process.

\begin{quote}
\textit{``First, the melody familiar to the visitor was made into a MIDI file, and then he was asked to create his accompaniment. AI can help him generate different harmonies and different accompaniment styles. He can choose from these musical materials according to his expectations or his mood.''} (P1)
\end{quote}

\textit{(2) Emotional Acceptance.} The therapy session will move on to the emotional acceptance stage once emotional awareness has been achieved. This will enable the client to understand their emotions objectively and make preparation for moving on to the leaving stage. Both cases mention music listening to help clients reflect on their current emotional state. In Case 1, \textbf{Emo-Gen} creates personalized emotional music based on the client's music preferences. In Case 2, improvisation
%singing 
refers to the request to clients to verbally compose 
%hum 
a melody representing their emotions, which can then be transformed into an instrumental piece using \textbf{Tone-Tran} techniques. The client listens to the melody with the therapist and objectively evaluates the emotions involved. This stage facilitates the client's acceptance of their own emotions.
\begin{quote}
\textit{``Ask him to hum a melody by himself and then turn his hummed melody into the sound of an instrument. That is, present the melody he hums in a new timbre. It might bring out a different feeling.''} (P1)
\end{quote}
%%就当下的一个语调来去哼一个自己想到的旋律或者自己创造的旋律。就是通过我刚所说的就是通过第一个方式就那个情绪就是情绪旋律生成之后，可以用这一个来去让他自己哼一段旋律，然后选择一个乐器，把他哼的这个旋律变换成那个乐器的声音也很好。对，就变成一个不同，就像刚刚子佳所说的就是另外一种音色来去表达它所呈现的东西，可能有不一样的感受。对。

%Therapists proposed implementing music tone conversion into improvisation, similar to the design above. 
This would allow the client to record and convert improvisational pieces into complete music. Clients can identify, accept, understand, and further reflect on their creations by listening to them.

%%Both Case 1 and Case 2 was targeted on understand and accept emotion. For Case 1 who has more difficulties expressing emotions, music therapists consider emotional melody generation to allow client rethink their emotions and accept them. 

\textit{(3) Emotional Coping.} In this stage, therapists focus on addressing negative emotions and the problems these emotions cause. When transitioning to the emotion leaving stage, participants consider the first step to cope with current emotions. Music therapists will help clients deal with existing emotional issues without generating new emotions. The approach of music listening reappears. However, in Case 1, the therapists aim to use \textbf{Emo-Gen} to alleviate the adverse effects of negative emotions experienced by the client rather than use emotional music to resonate her feelings. As P3 mentioned:

\begin{quote}
\textit{``Music with low arousal and positive valence is needed to help the client improve sleep.''} (P3)
\end{quote}

For Case 2, therapists believe that it is necessary to calm the client first, as he may be experiencing anxiety and high emotional arousal. Reducing arousal can help soothe the client and gradually generate more positive emotions. %Therapists chose the ISO music listening method, involving playing music related to the client's emotions and gradually transitioning to more relaxing music to reduce the high level of arousal and promote relaxation. 
Therefore, the therapist opted for the ISO principal Music Listening Method, which involves initially listening to music that matches the patient's current mood and then gradually transitioning to listening to music that represents the desired emotion. In this study, the therapist chose this approach to reduce the emotional arousal level of case 2 and promote relaxation~\cite{starcke2021emotion}.

\begin{quote}
\textit{``Unlike emotion in Case 1, he is more agitated. It's best to ask him to listen to a piece of music that matches his emotion and then (use AI to) gradually change the music style to bring his emotion down to a level we want before moving on to the later work.''} (P4)
\end{quote}

Additionally, \textbf{Emo-Gen} can generate music that resonates with clients' personal feelings and support refined adjustments and tuning.

\textit{(4) Emotional Transformation.} This is the final stage of releasing emotions. Music therapists assist their clients in transforming into the desired positive emotion. Depending on the client's cognitive perception, this transformation may occur after or during emotional coping.

The music listening method could be used to support Case 1 in developing positive emotions. The \textbf{Tone-Tran} technique provides clients with new feelings based on existing music pieces. Reproducing familiar music with variations could help clients develop new emotional experiences.

\begin{quote}
\textit{``Firstly, the music that the client likes is used to help the client quickly integrate into the treatment, and then the tone transformation is used to bring new feelings, break the inherent impression, and help them experience new emotions.''} (P3)
\end{quote}

Furthermore, therapists believe that \textbf{Emo-Gen} can gradually direct the client's emotional state towards the target mood. While music therapists usually achieve this through live performance, novice therapists may find it challenging to control music with clients' dynamic emotions due to their lack of experience. Therefore, musical AIs can assist therapists by generating music that matches the client's changing emotions and enables quick changes and adjustments. To ensure smooth emotional transitions between pieces, therapists suggested using \textbf{Genre-Tran} to create transitions between the generated pieces.

In addition, therapists chose recreation at the emotional transformation stage. Algorithms for \textbf{Melo-Har} (for harmony and accompaniment generation) can reduce clients' resistance during treatment and lower the threshold of music creation. Clients could achieve positive emotions by recreating music with the help of AI.

\begin{quote}
\textit{``Firstly, different harmonies are generated for the music, and then the clients are invited to accompany the music according to the generated harmonies to form their music. This method can help them rebuild their positive attitude and build self-confidence.''} (P5)
\end{quote}

The majority of therapists recommended utilizing \textbf{Tone-Tran} in Case 2 for clients with limited music experience. Their initial suggestion involves inviting clients to hum their emotions and transforming the sound into other instruments. This creation is also a form of improvisation. Following this, the music therapist will encourage the client to listen to this music and prompt them to develop it further. The tone transfer technique has the potential to produce harmony and accompaniment to the client's hummed melody, ultimately leading to the generation of new music and emotions.

\begin{quote}
\textit{``Save the melody that they hum themselves as a melody in another timbre, and let them make the harmonies and accompaniment for this melody by themselves, turning it into a new musical fragment. In this way, they may develop a sense of success because the whole therapeutic model is success-oriented. At the same time, there is a new feeling when they hear what they have expressed being presented as a complete piece of music.''} (P8)
\end{quote}

\subsection{Potential Benefits of Musical AIs on Music Therapy}

\begin{figure}[h]
\centering
\includegraphics[width=1\columnwidth]{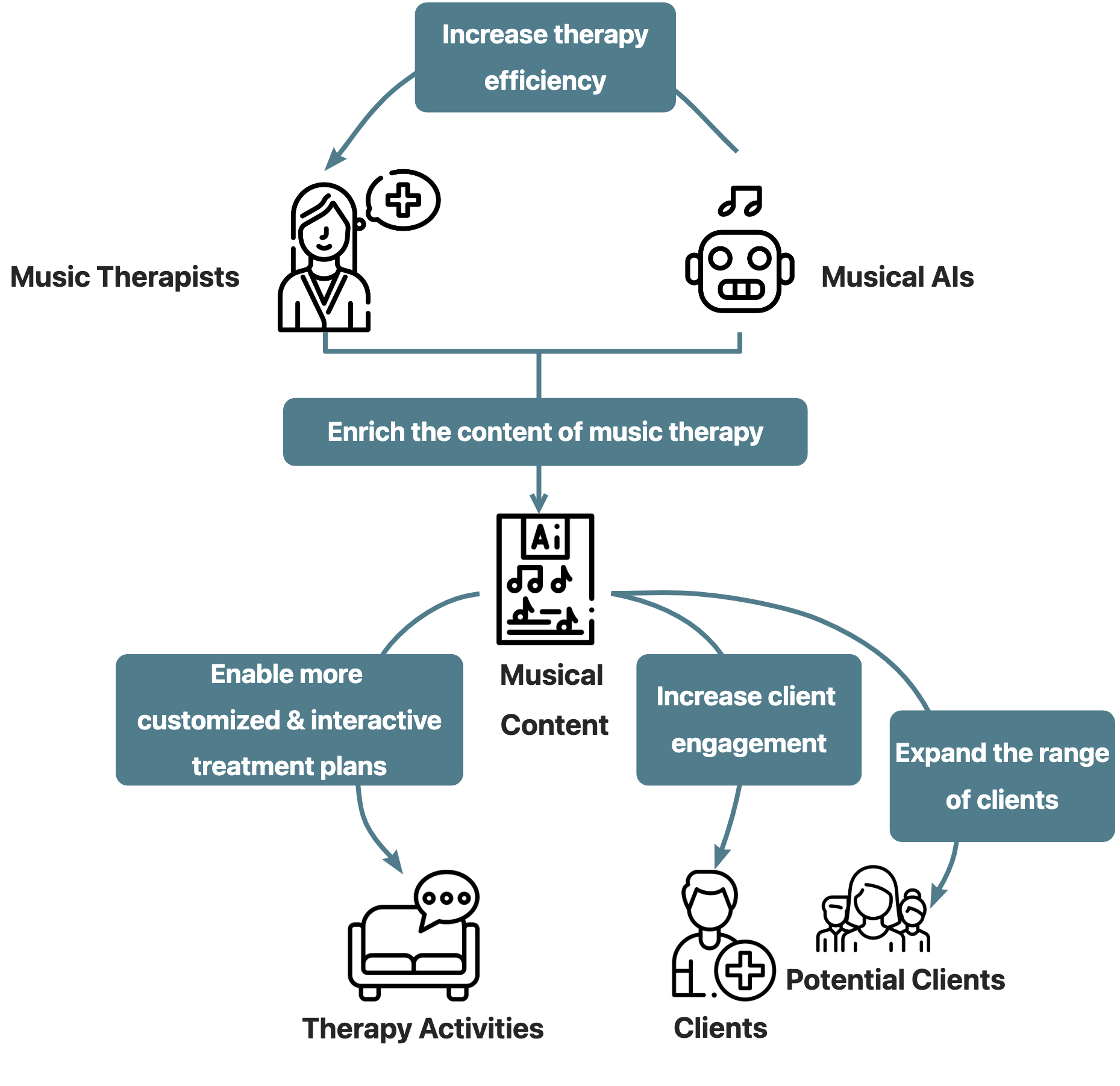}
  \caption{\textbf{Potential benefits of musical AIs on music therapy} }
    \label{fig: potential benefits_therapy}
    \Description{Figure 3 shows the potential benefits of musical AIs. These benefits are delineated across three dimensions: from the therapist's perspective, Musical AIs enhance therapy efficiency; within the context of therapeutic activities, Musical AIs have the potential to diversify therapeutic plans. From the client's standpoint, it may increase their engagement and broaden the range of clients. Overall, Musical AIs enrich the content of music therapy. This visual representation encapsulates these prospective benefits.}
\end{figure}

Gathering participant feedback and opinions has revealed several potential benefits of integrating musical AI into music therapy. These advantages encompass enhanced therapy efficiency, enriched therapy content, personalized treatment plans, and boosted client engagement alongside an expanded client base.

\subsubsection{Increase therapy efficiency} Participants believe that musical AI can assist music therapists in efficiently performing and producing music-related tasks and simulating musical instruments when they are impractical or unavailable. Therapists suggest that music generation techniques, namely \textbf{Ge-Gen} and \textbf{Emo-Gen}, could lighten their workload by aiding in composition and generating harmony and transitions between musical pieces. Participants emphasized the need for consistency and uninterrupted sessions in music therapy, noting that pauses could hinder clients' engagement with their emotional states.

\begin{quote}
\textit{``If the technique is suitable for use, I just need to select two pieces of music and generate the music for transition, which includes harmony. If all transitions can be generated, we don't need to manually create the transition using another piece of music (by ourselves). This could be more efficient.''} (P4)
\end{quote}

Furthermore, participants believe the \textbf{Tone-Tran} technique could lessen the need to transport numerous instruments for client outreach services or community programs, proving especially useful in group sessions requiring multiple instruments. P4 elaborated on this benefit:

\begin{quote}
\textit{``When preparing for group activities, we cannot prepare instruments comprehensively, especially if they are not portable, like drums, cellos, etc. Sometimes, we can only use the guitar. The AI technique can generate the tone of a desired musical instrument when it is not available in the client's place.''} (P4)
\end{quote}

Participants anticipate that musical AIs will facilitate music-making and editing tasks, enabling therapists to focus more on monitoring their clients' emotional shifts.

\subsubsection{Enrich the content of music therapy}
% Through the categorization and analysis of the AI opportunity in music therapy, it clearly emerged that the music therapists considered their current music library is not enough. 
Participants consider that musical AIs have the potential to expand the musical content available for music therapists to use during treatment. These techniques could broaden music selection options, inspire therapists during music creation, and facilitate playing unfamiliar genres or instruments. When creating a music library, participants mentioned that they either use music that is referenced from educational materials, use the music library curated based on peer review among coworker music therapists, or self-created playlists based on subjective categorization. However, they noted an issue: music libraries often don't meet clients' diverse needs. Thus, therapists expect that musical AIs could support generating certain types of music to enrich the current playlists for receptive music therapy. As mentioned by P7:
% For instance, there are typical songs that could represent people and reflect certain generations. Participants expected AI could generate personalized songs that are suitable for treating people from a certain generation.
% \begin{quote}
% \textit{"I hope when treating clients born in the 60s and 70s, the AI could generate some music songs related to Deng Lijun[a Taiwanese singer] because there were only a few songs in the previous library and it was really not enough. Sometimes I want to expand it. It would be great If it could generate songs related to certain groups of people during the treatment." (P7)}
% \end{quote}
\begin{quote}
\textit{``Sometimes I want to expand my music library. It would be great If it could generate songs related to certain groups of people during the treatment.''}~(P7)
\end{quote}

P1 suggested that music randomly generated by \textbf{Ge-Gen} could potentially inspire music therapists. Additionally, junior music therapists practicing improvisation may find AI-generated music a valuable reference for self-training.

\begin{quote}
\textit{``I'm currently treating different types of clients, I need lots of different harmony directions. The randomly generated music could give us more inspiration...
The junior therapists can use \textbf{Ge-Gen} as a reference during the learning process.''}~(P1)
\end{quote}

Therapists mentioned that AI-generated music could compensate for their limitations in musical performance abilities, including a perceived lack of experience producing certain musical instruments or genres.

\begin{quote}
\textit{``It is impossible for every therapist to be good at all the instruments and music styles. For example, one of my clients likes saxophone, but I don't know how to play it. I can use \textbf{Tone-Tran} to present its tone. e.g., when the client hums 'de de de' (simulating human singing), then AI can transfer it to the tone of a saxophone.''}~(P1)
\end{quote}

\subsubsection{Enable more customized content for treatment.} 

Due to individual differences and emotional complexity, customized treatment is demanding and challenging~\cite{baglione2021understanding} in music therapy. Understanding and accurately describing clients' emotions is a key objective for music therapists, as it significantly improves the quality of the treatment. Participants expressed interest in using musical AIs to enhance their empathy skills in the workshops.

 P1 made an example of generating music tailored to clients' personal experiences and preferences that might facilitate the therapy:
 
 \begin{quote}
 \textit{``We can use the client's favorite music as a starting point to let them generate their own harmonies and accompaniments that match their life experience. In this way, music can integrate personal experiences and scenes, like fragments. The client can combine the emotion with specific scenes to make music that fits their own story.''} (P1) 
 \end{quote}
 
P3 expects that musical AI could combine the current music therapist's library, the client's current mood, and the event context:
 \begin{quote}\textit{``Currently, the music we use is often sourced from pre-organized music libraries. However, if an AI could generate music  matching the client's mood and the background of an event, it would be truly unique.''} (P3)\end{quote}
 
P4 hopes that musical AI can generate music stories rooted in clients' personal experiences, evoking deep resonance with their personal narratives~\cite{cai2023listen}:
\begin{quote}\textit{``You can use a client's own personal experience to generate a very personal story about the music so the client can be more easily captivated by it.''} (P4)\end{quote}

\subsubsection{Increase client engagement.}
The concept of engagement has been defined as associated with cognitive, emotional, and behavioral components~\cite{kahn1990psychological}. Cognitive engagement refers to the level of concentration and involvement during activity participation, emotional engagement focuses on the individual's feelings and emotions, and behavioral engagement refers to the actual action. We will introduce how musical AIs increase client engagement from these three aspects.

\textbf{\textit{Cognitive engagement.}} Integrating musical AIs with music therapy theories can enhance music creation and facilitate the elicitation of diverse emotions in clients. Participants also consider that musical AIs may be able to assist clients in \textit{controlling their attention, thus better expressing and shifting emotions}. Additionally, music therapists hope AI-generated music could be more creatively utilized to help clients find inspiration and enhance their ability to express emotions. Participants would consider AI capable of creating music that represents their feelings by analyzing their verbal expressions, further contributing to the therapeutic effect. P4 explained:
\begin{quote}
\textit{" I can see that this (AI) technique can be applied in either improvisation or creation. Tone transformation (\textbf{Tone-Tran}) can be used to express different emotions and (can help clients) become aware of the current emotion. In Case 2, the client can just hum a melody that comes to mind (then he can engage and be aware of his inner emotions). 
"} (P4)
\end{quote}

\textbf{\textit{Behavioral engagement.}} Participants believe musical AIs can help clients express their inner emotions, especially those who may be hesitant, as indicated by P1:

\begin{quote}
\textit{``In Case 1, the client is more introverted and may not know how to express herself, but she is familiar with the music style and playing melodies. So I will ask her, 'What song do you like that can resonate with your feelings more?' Then, she may have chosen a certain song. Then, I will tell her that the song would be adapted later. When she changed the song, what she changed (the music style) may be unexpected. Through her change,  I will further let her try music imagery based on the music she just created. This process might trigger her to make some associations through the music and learn more about her inner self.''} (P1)
\end{quote}

P3 mentioned that AI-generated music may eliminate clients' fear of making music.

\begin{quote}
\textit{``With the help of AI, user can create melodies without the fear as the music can be generated with just 'one-click'.''} (P3)
\end{quote}
    
\textbf{\textit{Emotional engagement.}} Participants noted that AI-generated music can introduce new emotions to clients and foster group discussion, enhancing socialization. P4 explains:
\begin{quote}
\textit{``
Let's say you switch the style of the music through AI. For example, you listen to the original music or familiar songs like Jasmine Flower (a Chinese folk song) at an event. Then you use the technology to develop a jazz version or a variety of styles of the ensemble version of this song, which may stimulate discussion about music and promote their socialization.'' 
} (P4)
\end{quote}

\subsubsection{Expand the range of clients}
Musical AI has the potential to alleviate challenges in playing or producing music, especially for those without a musical background.
This capability allows individuals lacking musical knowledge to participate in music therapy sessions, encouraging their involvement without intimidation. Moreover, music therapists believe that enhancing clients' musical skills enables participation in a wider range of musical activities. 

P4 and P7 noted that AI-generated music could potentially support individuals with poor emotional well-being who might not require formal therapy.
\begin{quote}
\textit{``For those in need of emotional support—such as individuals with health issues, young people stressed from work, or the elderly—music activities can offer solace. Although these groups differ from typical clients, the same musical AI can support various activities for diverse audiences.''} (P7)
\end{quote}

\subsection{Concerns about Applying Musical AIs to Music Therapy}

Participating therapists have expressed concerns about human-AI collaboration in the context of music therapy despite the potential benefits of musical AIs. Some of these concerns arise from the professional role of therapists as healthcare providers and the unique characteristics of music therapy. Another factor is the scarcity of proficient musical AIs and associated application systems.

\begin{figure}[h]
\centering
\includegraphics[width=1\columnwidth]{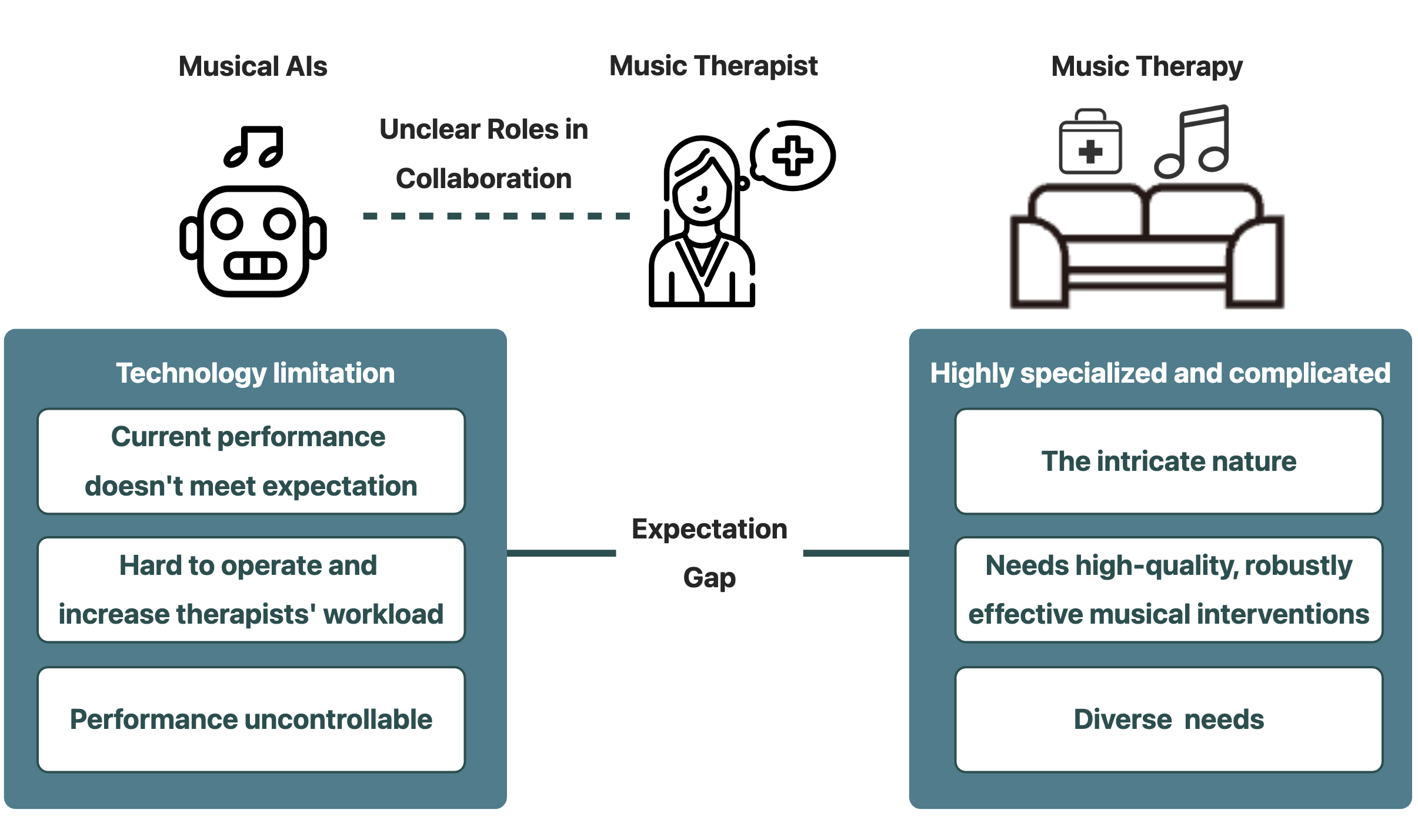}
  \caption{\textbf{Concerns about musical AIs in music therapy} }
    \label{fig:Concerns}
    \Description{Figure 4 presents concerns about applying Musical AIs to Music Therapy. Specifically, Musical AIs exhibit several technical limitations. Their outcomes may not align with human expectations. Moreover, music therapy is a demanding and specialized field characterized by a complex process, necessitating the use of high-quality and diverse musical materials. And the relationship between music therapists and AI remains unclear. This figure illustrates the gap in expectations between music therapy and existing Musical AIs.}
\end{figure}

\subsubsection{Music therapy is specialized and complicated}

Nearly all participants contend that music therapy entails a far more intricate therapeutic course than the mere playing of melodious compositions. Their professional endeavors encompass a profound comprehension of the client's psychological and emotional intricacies, coupled with the formulation of therapeutic objectives and strategies predicated upon medical acumen and experiential wisdom. Furthermore, it's important to recognize that the therapeutic modalities adopted by each music therapist are inherently idiosyncratic, adaptable, diverse, and improvisational, thereby rejecting a standardized paradigm. The detailed opinions are shown as follows.

\textbf{(1) The intricate nature of music therapy.}
%Even if musical AIs are expanded into music activities, that is still not enough.
Some participants mentioned that the intricate nature of music therapy could present challenges for adopting musical AIs. While musical AIs can enhance musical activities, they might not be adequate to fulfill the complex demands of music therapy. This is because music therapy encompasses more than just listening to pleasant music; it requires music therapists' expert knowledge and clinical experience.

P6 mentioned that:
\begin{quote}
\textit{``For a music therapist, clarifying the design goal is paramount. It has relatively high comprehensive requirements. Firstly, musical literacy and the comprehensive quality of capturing the other party's emotional state and understanding their psychology.''} (P6) 
\end{quote}
%% 我觉得设计治疗活动不是很重要的一个部分，因为只要你把案例理清楚，你的治疗目标梳理清楚，有无数种活动你都可以使用在后面
%%它的综合要求是比较高的。首先音乐素养，还有是就是捕捉对方情绪情感和掌握对方心理的这种综合素质综合要求

Additionally, P5 highlighted the complexity of the training required for music therapists:
\begin{quote}
\textit{``It requires a high level of training to be a music therapist. First, you need to understand music aesthetics, and then develop sensitivity to capture the client's emotions and understand them in psychological ways.''} (P5)
\end{quote}
% Cultural difference is another area that participants think AI is hard to adapt to. P3:
% \begin{quote}
% \textit{ "Which cultural music psychology is the AI based on?"} 
% \end{quote}

\textbf{(2) The high requirement for quality and robustness.} Music therapy requires high-quality, effective musical interventions, whereas using uncertain musical AI introduces significant risks.
Music therapy typically focuses on clients dealing with negative emotional states, emphasizing the importance of effectively addressing their emotional challenges. However, the current state of musical AI technology has not attained levels of efficacy that would satisfy music therapists. As a result, participating therapists are concerned that experimenting with immature techniques could have serious consequences. Several participants have mentioned not being satisfied with the results generated by the music in our workshop. For example, P1 points out that:

\begin{quote}
\textit{``The effect is crucial. Even if the music is thematically interesting, they will still resist if they dislike the sound.''}~(P1)
\end{quote}

\textbf{(3) The diverse needs of music therapy.} Meeting all these diverse needs with a single AI algorithm poses a challenge.
Music therapy is person-centered, customized care that is highly flexible and dynamic. 
%P2 explained: 
%\begin{quote}
%\textit{"This is a probability question. Maybe lots of people like this music, but based on what? Its music has a background. The background will contain a sort of emotion, it is personalized. The music that clients listened to from their childhood can trigger more [emotions]."}
%\end{quote}
Our participating therapists pointed out that different music therapists have different needs, which may challenge the development of musical AIs. For example, P3 points out:
\begin{quote}
\textit{``For instance, different therapists have different needs. If some therapists are majoring in piano, they do not need to use musical AIs to support musical performances. For others (who are not majoring in piano), sometimes they have to use the piano for a special client, but their technical skills may not allow them to implement the assumption therapy plan (in piano). They need some technology to help them.''} (P3)
\end{quote}
%% 比如说一个治疗师对每个人的需求不一样，有的治疗师他是钢琴专业毕业的，那么他在用钢琴这个乐器来激竞演奏或者说去创作的时候，其实他是很得心应手的，他没有必要去借助 AI 音乐的帮助来实现他的治疗设想不是钢琴专业毕业的，那有时候我的设想可能我的技术没有办法去实现。

\subsubsection{Musical AIs' technology limitations.}
The wave of deep learning advancements has fueled the development of musical AI. The participating therapists think that the technology's current capabilities are still limited after taking part in it and having firsthand experience.

\textbf{(1) Underperforming musical AIs.} Current performance of musical AIs' doesn't meet expectations.
While music therapists acknowledge the pleasantly surprising advancements in musical AI technology, they contend that the current efficacy of AI-generated music remains rather restricted, particularly in terms of emotional expression.
%\begin{quote}
%\textit{"In fact, I can still hear very obvious sounds of the machine".}~(P3)
%\end{quote}
Participants are more concerned about how musical AIs match emotions, resonate with clients, and are able to capture delicate emotions. P3:
\begin{quote}
\textit{``Perhaps I have high expectations. I feel that if I had listened to the generated music from four different emotion categories and then chosen the corresponding four emotions, I may not have chosen right.''} (P3) 
\end{quote}

P5 criticized that musical AI is not able to match delicate emotions,
\begin{quote}
\textit{``The main challenge is that delicate emotions are hard to capture, especially using the machine. It does not have temperature and no emotion. If you use rigid technology, even if you can operate it, it is still made of a bunch of mathematical formulas and models. It (the generated sounds)  is certainly not as good as music produced by humans.''} (P5)
\end{quote}
%P8 felt that generated music is hard to match with emotions, 
% \begin{quote}
% \textit{"Since emotions themselves are hard to define. We need to very carefully detect the emotional stage of the clients or clients, which may not be categorized as one or two options. In this respect, we still need a better integrate the assessment test and enhance the selection mechanism. I do not know how to make the selection yet. Maybe providing more options to let them [clients] select. Just like the emotional wheel with detailed categorizations for emotions, or maybe just let the clients answer a series of questions to describe their emotions."}
%\end{quote}

\textbf{(2) Low operability of musical AIs.} Musical AIs are hard to operate and increase therapists' workload.
Moreover, participants doubted how to effectively integrate musical AIs without disrupting the coherence of therapy.
%\subsubsection{Add unnecessary burden to the therapy process}
Some participants mentioned that musical AIs may add unnecessary workload to the treatment process:
 \begin{quote}
\textit{``Let's say playing guitar. I only need 30 seconds to prepare, but I have to set up the computer when using AI, despite picking up my guitar. The whole process increased my workload. So what is the point of me using it?''}~(P6)
\end{quote}
 \begin{quote}
\textit{``When my clients come, I cannot let them wait and adjust parameters on a computer and then let them listen to the generated music. It is hard to integrate into my therapy sessions.''}~(P3)
\end{quote}

Besides, our participants point out that there is a lack of applications and friendly interfaces for music therapists:
%Last but not least, participants lacked confidence about the technique's effectiveness when put into practice. 
\begin{quote}
\textit{``It seems there is a huge gap between hypothetical techniques and therapy. Since there is no user interface at the moment, so many things are unknown [...] I think at least I am not ready to say that I can take it (musical AI) to real clinical applications. Or do I still need to learn more about using it before I can use it in the clinic?''}~(P3)
\end{quote}
P5 also mentioned that she held a vague feeling about how to use AI technology:    
\begin{quote}
\textit{``I am unsure whether it would be made into an app or something else. I want to know what form it will be to know where to apply it. For example, I may use a certain treatment plan or client type. For now, if you want me to replace a certain [treatment] part with the machine, I feel many places do not go smoothly. I don't know how it connects with the interface.''} (P5)
\end{quote}

Participants raised several concerns about employing AI techniques. They voiced fears of increased workload for operations and the need for more effort to learn new knowledge. Additionally, they noted the inconvenience of pausing sessions for technique adjustments and expressed uncertainty regarding the usability implications of adopting the new technology.

\textbf{(3) Uncontrollable performance of musical AIs.}
Musical AIs have been observed to have unpredictable performance. Concerns have been raised by some participants about the potential negative impact of the generated sound on listeners or clients. P4 expressed her concern about randomly generated music:
\begin{quote}
\textit{``How can I guarantee at least 99 percent of the randomly generated music will not negatively impact my clients? I am afraid it will be under control if it (musical AI) irritates clients. If not, I will worry about the client-provider dispute. Of course, there is no conflict with AI technology itself, just that the impacts on clients are not controllable.''} (P4)
\end{quote}

%P5:
%\begin{quote}
%\textit{"A detailed assessment of the client is needed before starting the treatment. AI brings so many uncertainties when implementing it into the therapy process. We need to constantly adjust our treatment plan to adapt to the clients based on their responses. The treatment is basically being disrupted if you spend too much time operating (the AI). We don't have that much time, and need to quickly conduct the treatment process. You cannot let me spend a lot of time doing the delicate operation on this [AI]. It cannot be too complicated."}
%\end{quote}

\begin{quote}
\textit{``I don't know how it works practically. Not sure [...] how different from our imagination. If that is too difficult to operate, or the generated sound is completely different from what we want to express, or clients hear the music and say they do not feel that way.''}~(P2)
\end{quote}

\subsubsection{The unclear role of therapists in human-AI collaboration.}
%\subsubsection{Hard to balance the relationship between AI and therapy}
%During the co-design workshop, we observed that, in comparison to other professions, music therapists, functioning as healthcare providers, exhibit a heightened sense of responsibility towards their clients, encompassing stringent self-expectations and a commitment to patient well-being.

%\underline{\textbf{}}

In addition to expressing worries about how to use musical AIs in therapy scenarios, participants also questioned their relationships with AI. In the discussions of P3 and P4, they considered their role in therapy~\textit{``becoming sort of software operator''} instead of the music therapist. Moreover, some participants are ambivalent about whether musical AIs replace therapist-client interaction.
 \begin{quote}
 \textit{``What is the meaning of holding activities if let the machine replace [the interaction]? It feels as if improvisation is done merely for its own sake. In fact, we use music to build connections, feeling each other's emotions during the improvisation. During improvisation, we make adjustments and strive to match and resonate with their emotions.''}~(P6)
  \end{quote}
  
  \begin{quote}
 \textit{``Even though I'm interested in AI technology and always think about how it can be integrated into music therapy. However, my learning experience taught me to focus more on human interaction. Therefore, I would avoid using certain technology like this. So I would avoid using certain technology like this.''} ~(P8)
 \end{quote}

%What's more, considering AI algorithms are trained based on a large number of data, P3 expressed his concern about ethical issues regarding the training data collection:
% \begin{quote}
% \textit{"Although we made great assumptions, I think there may be ethical issues. For example, you have clients coming into your treatment room, and you not only treat them, but you take their responses to train the [AI] model. I don't think that is appropriate."}
% \end{quote}

\section{Discussion}

\subsection{Designing Musical AIs for Emotion-focused Music Therapy}
For the first step in exploring how to apply musical AIs in music therapy, we focused on solving a specific class of problems, which is targeted to benefit further system design in emotion-related music-therapist-AI collaboration. %Therefore, our topic focuses on music therapy interventions for emotional issues.

Therefore, we collaborated with music therapists to develop a comprehensive music therapy treatment plan, encompassing a typical music therapy workflow and an exploration of relevant musical AIs to support each stage of the therapy process for specific pre-designed cases. Discussions with therapists found three main therapy stages and four processes. The five relatively mature musical AI techniques mentioned previously, \textbf{Ge-Gen}, \textbf{Emo-Gen}, \textbf{Melo-Har}, \textbf{Genre-Tran}, and \textbf{Tone-Tran}, were presented to the therapists in advance and assigned to various processes of the therapy workflow during their discussions.
For example, in emotion awareness, therapists largely agreed that \textbf{Genre-Tran}, \textbf{Ge-Gen}, \textbf{Emo-Gen}, and \textbf{Melo-Har} can be used for music imaginary, improvisation, music listening, and recreation.

%We collected therapists' perspectives on musical AI techniques and synthesized a workflow for emotion-focused music therapy within the topic of music therapy for emotional problems.
To better organize the co-design results, we utilized a theoretical framework~\cite{greenberg2004emotion} for emotion-focused therapy as a typical workflow to frame the design suggestions from our workshop.
This workflow aims to aid researchers with limited knowledge of music therapy in linking musical AIs to specific tasks within the practice of music therapy. 
Furthermore, our investigation into musical AIs has enabled us to identify a current gap in using music AIs within music therapy. Musical AIs primarily concentrate on algorithm-based research, with minimal attention given to their application within therapy. Integrating these discrete technologies in a manner that supports various tasks performed by therapists and clients is crucial for their effectiveness within the context of music therapy.
%These music therapy methods in the emotion awareness stage can serve to evoke, experience, and express emotions~\cite{stewart2019music, mcpherson2014role}. 
This workflow aids us in pinpointing the intersections of music therapists and AI collaboration in addressing emotional issues.
We anticipate that AI technologies can serve as assistants to therapists, helping them achieve more effective intervention methods~\cite{novelli2022deep}. 

%We hope that our research findings will offer guidance and constructive suggestions for further design research in this area.

%%通过co design的发现,对人群的了解和讨论； 的设计,还需要adept到不同的case

In short, our research serves as an initial step in exploring the integration of AI into music therapeutic processes. Through this co-design workshop, we established a connection between specific music AI technologies and concrete therapeutic processes and methods. We believe this paper will shed light on further AI-supported music therapy research.
%the results of our two case studies show that musical AIs could support music therapy in various aspects, including working efficiency, therapy content, treatment plans, and client engagement. Nevertheless, therapists thought the relationship between therapists and AI was still unclear in human-AI collaboration, and the complexity of music therapy and the technical limitations of musical AI may also inhibit the adoption of musical AIs in therapy. We will discuss our findings and design implications for musical AI-based therapy tools.

\subsection{Potential Benefits of Musical AIs from Music Therapists' Perspective}

% The work of the music therapist directly influences the effectiveness of music therapy~\cite{lathom1982survey,luborsky1985therapist}. Therefore, we gathered the diverse perspectives of music therapists to gain insight into applying AI technologies to support music therapy. 

Overall, all of our participants agree that musical AI tools are advantageous for therapists, clients, and treatment outcomes. Music therapy is, in fact, a highly labor-intensive job. Music therapists are also at risk for burnout~\cite{gooding2019burnout}. Therefore, they appreciate that AI could support some repetitive and tedious tasks, for example, providing examples of transition chords, as mentioned by P1. Besides, the collaborative relationship between therapists and clients and client engagement influence the ultimate therapeutic outcomes~\cite{hill2005therapist}. Therapists are particularly interested in musical AIs that could contribute to client engagement in music therapy. For example, therapists thought that the enriched content of music therapy could increase client engagement.

To our surprise, although most music therapists are professionally trained, they still face challenges in therapy~\cite{rushing2017obtaining}. For instance, music therapists may encounter situations where they lack familiarity with a client's preferred music style or instrument. In such cases, the utilization of musical AI holds promise in enhancing their proficiency in executing various musical tasks. Considering the assorted skill sets required of music therapists, therapists expect musical AIs to assist them with different capabilities in therapy. Therefore, the design of musical AI tools can be tailored to the musical sophistication and competence level of therapists. 
Designing musical AIs to increase therapists' competence is a practice of developing hybrid intelligence to perform challenging tasks or solve complex problems~\cite{carter2017using}.

It is essential to emphasize that an effective assistive tool benefits the therapist and serves as a protective factor for treatment outcomes. In general psychotherapy, the therapist's variables directly influence the effectiveness of therapy~\cite{lingiardi2018therapists}. Notably, the effectiveness of treatment outcomes is positively correlated with therapist job satisfaction and inversely correlated with levels of burnout~\cite{delgadillo2018associations}. Additionally, treatment outcomes are also influenced by other work-related attitudes of therapists~\cite{sandell2007therapist}, including their confidence and enjoyment of their work. The therapist's skills and work attitude are directly linked to treatment outcomes. One significant reason for this connection is that the therapist's subjective factors can affect the establishment of a therapeutic alliance between the therapists and their clients~\cite{ackerman2003review}, which is a crucial predictor of treatment outcomes. Moreover, from the clients' perspective, AI could encourage clients to be more engaged in therapy. For example, clients may be reluctant to participate in a music therapy program due to their limited competence in music performance, while AI could lead them to finish some easy tasks at the beginning of music therapy or assist them in performing some challenging activities. Ultimately, AI techniques could encourage clients to participate more actively in therapeutic activities, thus benefiting more from the therapy~\cite{hill2005therapist}.

\subsection{Challenges in Human-AI Collaboration in Music Therapy}

While our participants generally perceive the introduction of musical AI as holding considerable potential benefits for music therapy, they have also articulated numerous concerns. We have observed that some of these concerns are common within the broader domain of human-AI collaboration.

Our participants point out that current musical AIs have some technical limitations that hinder their ability to provide accurate emotional music for effective therapy interventions, echoing previous work in human-AI collaboration~\cite{wang2021brilliant,cai2019human,chu2023work}. For example, when doctors collaborate with clinical decision-making systems, they find that the top recommendation provided by AI is not always right. In our case, generative AI's effectiveness, usability, and technological limitations in musical AI are notably pronounced, demanding significant further improvement. It is well-recognized that we should improve the algorithm and augment the related training datasets for musical AIs in the future.
Furthermore, the absence of relevant applications and interactive interfaces significantly hinders therapist-AI collaboration. Designers need to account for the specific requirements of the context and develop corresponding AI systems, which is a pivotal concern in the field of human-AI collaboration. For instance, prior research~\cite{jeon2021fashionq} has revealed that uncertainty regarding AI's capabilities and the complexity of AI outputs make human-AI interaction particularly challenging to design. Therefore, one of the future research challenges lies in devising user-friendly and comprehensible interaction interfaces for therapists, considering the technical limitations of musical AIs.
Additionally, the relationship between music therapists and AI in collaborative settings constitutes another focal point of research within the realm of human-AI collaboration. Prior studies~\cite{chan2022investigating,oh2018lead} have indicated that in collaborative endeavors with AI, designers may experience reduced agency and expressiveness as AI assumes some of their tasks, thereby potentially diminishing their sense of ownership. Hence, an essential future research direction pertains to establishing clear role delineation and task allocation between therapists and AI to enable both to leverage their strengths, safeguarding therapists' sense of control over the therapeutic process.

However, in addition to the challenges echoed in previous human-AI collaboration research, we have identified certain challenges that are unique to the context of music therapy and distinct from the broader domain of human-AI collaboration. Similar to other forms of art therapy~\cite{case2014handbook}, music therapy encompasses the seriousness and critical requirements of therapeutic contexts while embracing \textit{a high degree of diversity and flexibility} in therapeutic workflows and methods~\cite{baker2010music,rolvsjord2005research}. Because music therapy is a complex and person-centered process, it implies that methods beneficial for some individuals may not necessarily be suitable for others~\cite{abrams2018understanding,noone2008developing}. Sometimes, a single music therapy approach may not adequately address clients' needs, so therapists also utilize a range of intervention techniques and diverse music therapy theories during treatment.
Moreover, the \textit{high flexibility} in treatment also posed significant challenges in designing human-AI collaboration in therapy. Our participants indicated that they do not strictly follow fixed rules or frameworks. Instead, they prefer to adapt treatment plans flexibly according to the client's requirements and current state, incorporating improvisation.

Given that music therapy is a complex, highly specialized process that necessitates stability, reliability, and adaptability to diverse therapist and client needs, we collaborated with music therapists to delineate their workflow based on emotional-focused therapy~\cite{greenberg2004emotion} for exploring human-AI collaboration at different stages. Therefore, our research findings provide valuable insights into developing human-AI collaboration for a workflow characterized by a high degree of flexibility and diversity within a specialized field like music therapy.

\subsection{Design Implications for Musical AIs in Music Therapy}
%%案例间对技术选择的区别；措辞上的clarify:呈现音乐治疗师的选择,更倾向于哪种（不能算是AI更适合？）对case study 的设计,还需要adept到不同的case;音乐治疗师提的和researcher的设想

Our findings focused on using musical AIs to build treatment solutions for solving anxiety and depression-related emotional problems. We offer the following design implications for future HCI research on applying musical AIs to music therapy:

%\textbf{Implication \#1: Support in creating personalized treatment plans.} Music therapists are expected to provide personalized treatment because clients are more likely to relax, release stress, and develop a positive mood if the selected music matches their preferences~\cite{mornhinweg1992effects,labbe2007coping}. To personalize therapy, therapists need to consider various personal characteristics of clients, such as personal growth experiences, cultural backgrounds, musical preferences, and personalities~\cite{goetz2018personalized}. We think preference elicitation and emotion detection are particularly important in generating personalized music that can resonate with clients' emotions~\cite{cai2023listen}. Developing a personalized music library also requires continuous evaluation and scoring of the generated music.

\subsubsection{Implication \#1: Support in creating personalized treatment content.}
From our findings, we can see music therapy practitioners are increasingly expected to deliver personalized interventions, which align with previous research that adapting the chosen musical elements with the client's preferences enhances the likelihood of relaxation, stress alleviation, and the elicitation of positive emotions~\cite{mornhinweg1992effects, labbe2007coping}. 
To facilitate personalized therapeutic interventions, therapists often need to consider various individual characteristics of clients, such as personal developmental experiences and cultural backgrounds.
For music therapists, an essential element in personalization is the individual's musical preferences~\cite{goetz2018personalized}.
Researchers have created a chatbot that provides personalized music recommendations and comments by adapting to users' music preferences and emotions~\cite{cai2023listen}. Both recommended music and music comments could resonate with listeners' emotions and support mental well-being~\cite{jin2023understanding}. This aspect is echoed in our study, where music therapists emphasize the significance of music selection in therapy, with clients' music preferences serving as a crucial reference. 
Even the most experienced music therapists may not be proficient in every music genre, but AI technologies can aid in integrating personalized music into their sessions.  
This, in turn, aids therapists in accomplishing their therapeutic goals more effectively. %However, concerns from therapists in the present research have been raised about the quality of AI-generated music. Consider that, future designs should also include continuous evaluation and rating of AI-generated music, to foster the ongoing development and application of AI technologies in the field of music therapy.
Drawing from these insights, we propose that AI tools applied in music therapy should incorporate a variety of personalized functions.  Primarily, such a tool should be capable of accepting raw data input from users and, based on this input, analyzing the user's preferences for musical materials. 
%Additionally, considering that clients often find it challenging to provide extensive raw materials, this implies a need for rapid analysis capabilities in the tool. Therefore, a tool with personalized therapeutic functionalities should not only possess generative features but also rapid analysis capabilities. Building upon automation, to enhance the quality and usability of the generated materials, the tool should include options for manual adjustments.   For instance, there should be channels available for therapists to modify, particularly in the dimensions of fundamental musical elements such as melody, chords, and tempo. 
Additionally, to enhance the personalization of therapy, clients should also be allowed to provide feedback on the tool or the process of its use.

\subsubsection{Implication \#2: Incorporate domain knowledge of music therapy.} %theory.}
Our findings showed that music therapists expected musical AIs to be designed in depth with the theory of music therapy to assist them in tasks that require professional knowledge but are repetitive and time-consuming. %For example, P1 expressed her need for AI to provide examples of transition chords to help her quickly calculate the chord required. Based on our findings, studies on automatic chord recognition~\cite{chen2012chord} and progression~\cite{wu2020jazz,brunner2017jambot} would consider exploring the design opportunities for implementation in music therapy. 
For example, the ISO principle is a classic principle in music therapy, which states that in music therapy, clients are initially encouraged to listen to music that resonates with their current emotions, even if those emotions are not what they desire. Subsequently, therapists gradually guide clients to shift their emotions toward the desired state by altering the content and emotional tone of the music~\cite{starcke2021emotion}. Taking this principle as an example, therapists are suggested to modify the content and emotional trajectory of the music frequently and sensitively throughout the therapeutic process.
Therefore, both music generation and music transition AI technologies need to be considered simultaneously. This can facilitate the simultaneous music generation and transition processes, assisting music therapists in achieving therapeutic goals within the ISO principle~\cite{qiu2023generated}. 
The Guided Imagery and Music (GIM) is another example. It is based on producing effects through visual imagination~\cite{beebe2009guided}. Therefore, music AI technology can moderately consider incorporating environmental sounds, such as the sound of flowing water or the wind and rustling leaves. This could function like a toolbox for storing musical materials to help therapists better utilize music to shape visual imagination. %For music therapy methods that mainly involve music creation or re-creation, AI technology needs to pay extra attention to operability. When adapting an existing piece of music, the process of re-creation is crucial to therapy; an operational platform that allows collaboration between therapists and clients is preferable to a fully automated feature. 
While the examples provided are brief, it's essential to emphasize that technologies closely aligned with music therapy theories and methods benefit therapists.

\subsubsection{Implication \#3: Support the music therapists' control in musical AIs.} 
Implementing AI techniques in high-stakes problems, domain experts are more concerned about implementing safety and user acceptance than AIs' creativity~\cite{singh2020current}. 
In our study, participating therapists are expected to check the AI-generated results and make revisions before implementing them into the clinical process. To support therapists' control in collaborating with AI, we suggest designers incorporate the explanation of the working process of musical AIs or justification of generated content~\cite{bryan2023exploring}. 
Furthermore, given the variation in therapists' competency and requirements, the user control in musical AIs could also adapt to some of the therapists' personalized needs~\cite{jin2020effects,millecamp2018controlling}. 
In light of this, we suggest that music AI tools should provide therapists with a real-time reassessment of the results. This reassessment involves an evaluation of music stability or quality. Furthermore, akin to the first recommendation, there should be channels available for manual modifications by therapists. These mechanisms serve to assist therapists in enhancing control over the music AI.

\subsubsection{Implication \#4: Support multi-stakeholder interaction and operation.}
%Music therapists expressed the desire for an interface that supports their interaction with clients while being therapist-operated. They prefer direct control over music-related parameters on the interface, following traditional music composing/editing tools. Additionally, therapists expect minimal effort in managing musical AIs to avoid disruptions during treatment. In contrast, a user-friendly design is crucial for clients using the same interface to facilitate their active involvement in music therapy and accommodate diverse expression methods. Considering clients with limited musical knowledge is especially important, ensuring the interface is not overly complicated.
Music therapy involves both therapists and clients. %The design of interfaces demands attention, considering the perspectives of both music therapists and clients. 
Our participant music therapists may need an interface that facilitates their interaction with clients, emphasizing professionalism. 
However, music therapists and clients often have different needs for interface design. 
On one hand, as music experts, music therapists prefer direct control over music-related parameters, akin to features found in traditional music composition and editing tools. 
On the other hand, clients, given their limited musical knowledge, mostly benefit from a simple, non-specialized interface. A simple and accessible interface encourages clients' engagement in therapy and avoids imposing additional cognitive load~\cite{guo2023touch,gong2018escape,jin2021oyaya}. 
This underscores the importance of future designs addressing diverse dimensions of interfaces to cater to the distinct needs of multi-stakeholders. 
This implies that AI tools applied in music therapy should have at least two kinds of user interfaces. 
Illustrating the process of transforming vocal humming into instrumental music, clients without musical background could also engage in the therapy. 
Therefore, for therapists, a professional interface should allow for the categorization of diverse instruments and even include adjustments for music reverb effects. On the other hand, for clients, intricate instrument selection and personal fine-tuning music effects may not be necessary. 

\subsection{Limitations and Future Work}

We point out three limitations in this study. First, the sample size was relatively small as we required participants to have clinical music therapy experience related to anxiety, stress, and depression. Ideally, they should have been currently engaged in relevant work. Therefore, among those who expressed interest, only a small number met these criteria, leading to a limited sample size. 
Secondly, our findings may be influenced by gender bias. However, this gender imbalance aligns with the existing demographic composition of music therapists~\cite{puhr2018current}, reflecting profession-related characteristics. Nevertheless, gender differences must still be acknowledged, and future research should include more male music therapists to address potential bias.
Thirdly, the study may not fully represent perspectives from therapists in other demographic groups. While most participants were Chinese, they were residing and practicing in various countries across Asia and North America. Therefore, conducting broader research that includes therapists from diverse backgrounds is crucial. Given these limitations, further discussion and refinement are necessary, and they will be the focus of future efforts. Additionally, future work will involve developing interface prototypes to test design implications and further explore the interaction and collaboration between music therapists and musical AIs.

\section{Conclusions}
In this paper, we conducted semi-structured interviews and co-design workshops with music therapists to understand the opportunities and challenges of integrating musical AIs' into the current music therapy treatment process. Our findings show that, from the perspective of therapists, there is potential to apply musical AIs to improve various aspects, such as therapy content richness, therapy efficiency, and client engagement. At the same time, we identified challenges when implementing musical AIs into therapy, such as the technical limitations of musical AIs and the complexity of music therapy. Furthermore, we discuss our findings through the lens of music therapists and HCI practitioners, and we provide four design implications reflecting on the design of human-AI collaboration in music therapy. We hope our work could stimulate future research on human-centered AI for therapy applications.

\begin{acks}
This project is supported by the National Natural Science Foundation Youth Fund 62202267.
\end{acks}

\bibliographystyle{ACM-Reference-Format}
\bibliography{main}

\clearpage
\begin{appendix}
\section{Pre-defined Cases Details}
The details of two pre-defined cases are listed in Fig. 5.
\begin{figure*}[]
\centering
\includegraphics[width=\textwidth]{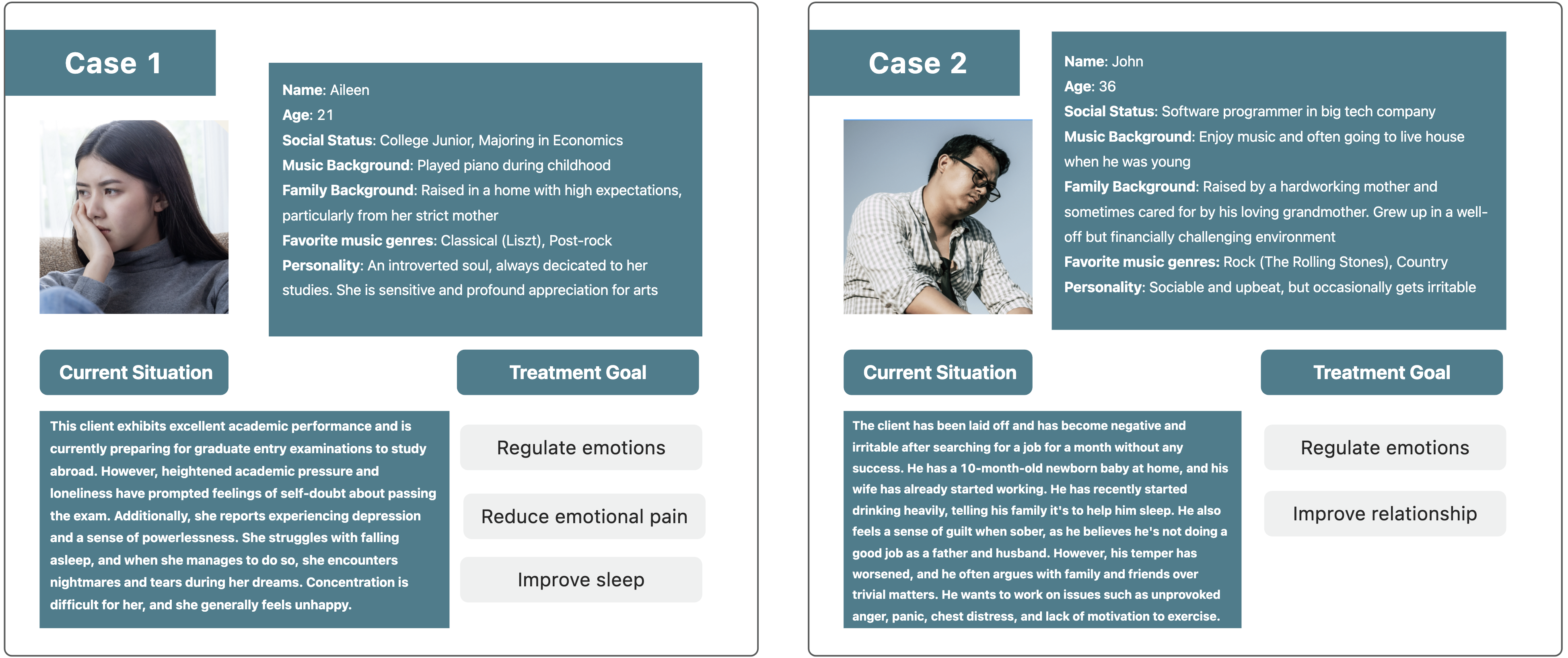}
  \caption{\textbf{Persona Design for Case 1 and Case 2} }
    \label{fig: case design persona}
\end{figure*}

\section{Typical Music Therapy Methods}
%In general, therapists are flexible in applying therapeutic techniques in therapy, which may be derived from different music therapy genres. There are many common music therapy methods and techniques, such as song sharing, song discussion, music listening, music imagery, music relaxation, music aesthetics, improvisation, and re-creative methods~\cite{wheeler2015music}. 
In our study, therapists mentioned similar ideas about using different methods and techniques in therapy. After a systematic summary, the therapists mentioned several methods that would be relevant to our research. So here are several methods that will be introduced for reference:

\textbf{Music listening} is one of the receptive methods that can help clients reduce stress and anxiety. Musical Sound Identity (ISO) is a principle of listening. It comes from Benenzon Music Therapy aims to match the client's inner emotions with the selected music by the therapist~\cite{benenzon2007benenzon}. During the process, therapists would first use music to empathize with the client's current mood and then change the music with different tempos and emotional state to gradually altering the mood and further guide the client to form new experiences.

\textbf{Music imagery} is another method similar to Guided Imagery and Music (GIM) created by Helen Bonny. This type of approach generally refers to listening to music accompanied by visual images (i.e., mental pictures). The imagery evoked by music is believed to increase self-awareness and relaxation, explore the subconscious, stimulate creativity, and assist in self-integration~\cite{wheeler2015music}.

\textbf{Improvisation} is often used in Creative music therapy or Nordoff-Robbins music therapy ~\cite{nordoff2007creative}. This method believes that interacting with music allows people to express themselves and form unique connections with others, thereby helping them feel a sense of self-transcendence or self-healing~\cite{aigen2014music}. Analytically Oriented Music Therapy also believes improvisation is an important intervention. But the difference is that this kind of intervention is related to Jungian Theory\cite {macdonald2013music}. It believes that music is a non-verbal symbol that can be used as material for psychoanalysis to help the therapist understand the client's inner world or to help the clients better express themselves\cite{eschen2002analytical}. And there are vocal improvisation and instrumental improvisation in music therapy.

\textbf{Re-creative} music therapy is another common therapy approach that refers to the process of recreating the original musical material through the vocal or instrumental music of the client to achieve therapeutic goals. Sometimes, these therapeutic methods manifest themselves as activities such as singing, musical games, or instrumental music playing. However, in the context of specific therapeutic goals, activities vary with specific techniques for achieving the therapeutic goals~\cite{wheeler2015music}.

\section{Semi-structured Interview Protocol}
\subsection{Demographic information}
\begin{itemize}
    \item [1.] Name:
    \item [2.] Gender:
    \item [3.] Age:
    \item [4.] Education background:
    \item [5.] Working Experiences:
\end{itemize}

\subsection{Interview Purpose}
\subsubsection{Main goals} Through interviews with music therapists, we aim to comprehensively understand music therapy theory and explore issues related to current research in this field.
\subsubsection{Sub-goals}
\begin{itemize}
    \item [1.] Be informed of the interviewee's current working process and current pain points;
    \item [2.] Ask detailed questions related to the research question;
    \item [3.] Prepare for the workshop setup.
\end{itemize}

\subsection{Interview Questions}
\subsubsection{Introduction of Experience}
\begin{itemize}
    \item [1.] "Can you describe your current research area and clinical experience?" (Based on the participants' background experience)
    \item [2.] "What is the main type of patient you work with?"
    \item [3.] "Under what circumstances is music therapy performed?"
    \item [4.] “Is there a specific problem that requires diagnosis, and to what extent can it be treated with music therapy?"
\end{itemize}

\subsubsection{Therapy Workflow}
\paragraph{Workflow Procedures / Steps}
\begin{itemize}
    \item  [1.] "Can you divide your workflow into stages or phases?" (Could you provide a brief introduction)
    \item  [2.] "Will there be any general principles and paradigms in the treatment process (such as for different treatment groups, ex. depression, or schizophrenia?)"
    \item  [3.] "How long is your single session?”
    \item  [4.] "What is the duration of the client's treatment cycle?"    
\end{itemize}

\paragraph{Question on Receptive Music Therapy}
\begin{itemize}
    \item [1.] "What types of music materials and electronic devices do you typically use when incorporating receptive therapy into your sessions?"
    \item [2.] "Do you have any experience building a music library that includes emotional songs before treatment? How do you determine which songs to include?"
    \item [3.] "How do you determine which songs correspond to different emotions?"
    \item [4.] "Will specific music be recommended to different patients?"
    \item [5.] "Will treatment strategies be modified to consider the patient's listening preferences?"
    \item [6.] "Would it be beneficial to know a patient's music preferences before starting their treatment?"
    \item [7.] "If you are able to generate music based on the patient's listening preferences, would you consider using it?"
    \item [8.] "Have you ever heard of AI-generated music?”
    \item [9.] "How would you use it? Would you prefer to listen to it in therapy, analyze it, or reference it and play it live?”
    \item [10.] "Do you have any concerns or worries about this?”
\end{itemize}
\paragraph{Question on Active Music Therapy}
\begin{itemize}
\item [1.] "Are there more active music therapy methods being used in individual treatment scenarios or group treatment scenarios compared to the receptive music therapy method in the clinic?"
\item [2.] "Do you typically use musical instruments or something else? If so, what kind of instruments?"
\item [3.] "What types of individual therapy (and group therapy) treatment are commonly used?"
\item [4.] "What is the duration and frequency of treatment?" (both individual and group treatment sessions)
\item [5.] "Can you share any specific examples?" (both individual and group treatment sessions)
\end{itemize}

\begin{figure*}
\centering
\includegraphics[width=\textwidth]{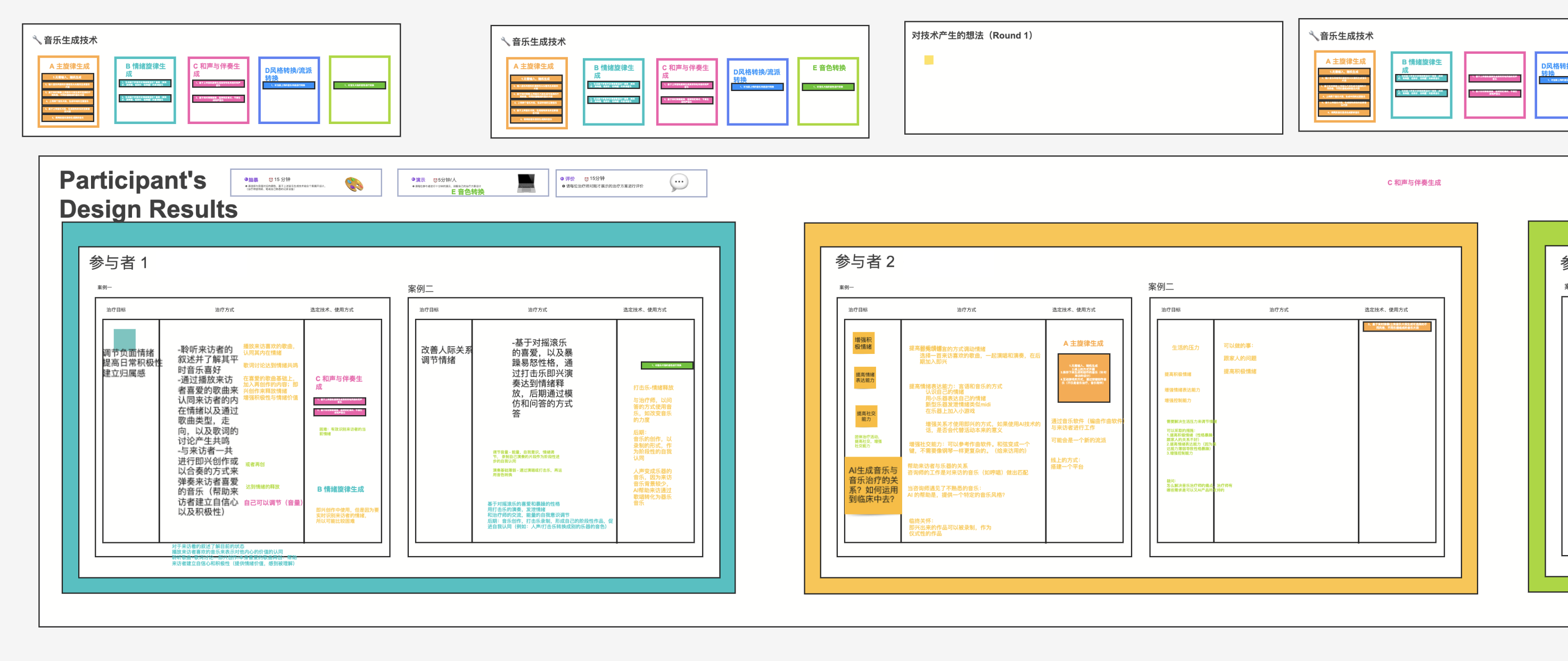}
  \caption{\textbf{Screenshot on Design Results} }
    \label{fig: indivisual design results}
\end{figure*}

\begin{figure*}
\centering
\includegraphics[width=\textwidth]{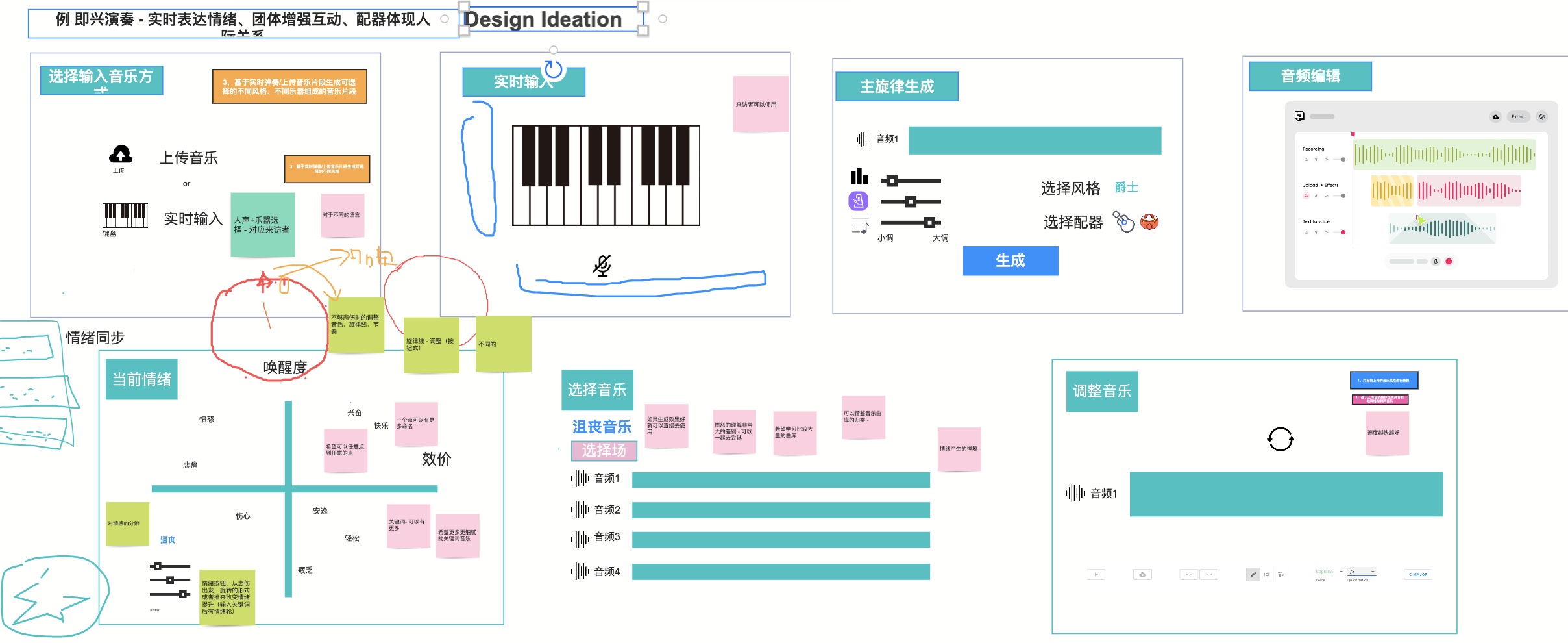}
  \caption{\textbf{Screenshot on Design Ideations} }
    \label{fig: indivisual design implementation}
\end{figure*}

\paragraph{Addition Questions}
\begin{itemize}
\item [1.] "What are your thoughts on the current improvement needs or pain points in this area?"
\item [2.] "Do you have any further questions?"
\end{itemize}

\section{Survey for AI-music Learning Materials}
\paragraph{instruction}
\paragraph{"Please complete this survey after you have finished learning. Thank you for your participation!"}
\paragraph{Questions}
\begin{itemize}
\item [1.] "What kinds of musical AI technologies are you interested in? " [Select multiple choices]
\begin{enumerate}
    \item [A.]General Melody Generation
    \item [B.]Emotion-based Music Generation
    \item [C.]Melody Harmonization
    \item [D.]Music Genre Transfer
    \item [E.]Tone Transfer
\end{enumerate}

\item [2.] "What features of musical AI technologies are most important to you?" [Open question]
\item [3.] "Do you have any questions or thoughts about the material that you would like us to know?" [Open question]
\end{itemize}

\section{Screenshots of Co-design Workshops}
The screenshots of our co-design workshops are listed in Fig.~\ref{fig: indivisual design results}, and Fig.~\ref{fig: indivisual design implementation}.

\end{appendix}

\end{document}